\documentclass[a4paper,12pt]{article}
\pdfoutput=1
\usepackage{graphicx,rotating,colortbl,hyperref,slashed,amsmath,xcolor,amssymb,amsfonts,cite,rotating}

\makeatletter
\hypersetup{colorlinks,bookmarksopen,bookmarksnumbered,
linkcolor=blus,pdfstartview=FitH,urlcolor=rossos,citecolor=verde}
\usepackage{lscape}

\def\lsim{\mathrel{\rlap{\lower3pt\hbox{\hskip0pt$\sim$}}
   \raise1pt\hbox{$<$}}}         
\def\gsim{\mathrel{\rlap{\lower4pt\hbox{\hskip1pt$\sim$}}
   \raise1pt\hbox{$>$}}}         

 \newcommand{\One}{1\hspace{-0.63ex}\hbox{I}}

\newcommand{\mio}[1]{}
\newcommand{\xxx}[1]{{\color{red}[\bf #1]}}

\definecolor{Gray}{gray}{0.95}

\newcommand{\X}{\digamma}
\newcommand{\Q}{{\cal Q}}

\newcommand{\fb}{\,{\rm fb}}
\newcommand{\NTC}{N_{\rm TC}}
\newcommand{\NF}{N_{\rm TF}}

\newcommand{\sfrac}[2]{#1/#2}

\usepackage{multicol}
\usepackage{color}
\definecolor{rosso}{cmyk}{0,1,1,0.4}
\definecolor{rossos}{cmyk}{0,1,1,0.55}
\definecolor{rossoc}{cmyk}{0,1,1,0.2}
\definecolor{blu}{cmyk}{1,1,0,0.3}
\definecolor{blus}{cmyk}{1,1,0,0.6}
\definecolor{bluc}{cmyk}{1,1,0,0.1}
\definecolor{verde}{cmyk}{0.92,0,0.59,0.25}
\definecolor{verdec}{cmyk}{0.92,0,0.59,0.15}
\definecolor{verdes}{cmyk}{0.92,0,0.59,0.4}

\newcommand{\eq}[1]{~{\rm (\ref{eq:#1})}}

\newcommand{\MeV}{\,{\rm MeV}}
\newcommand{\GeV}{\,{\rm GeV}}
\newcommand{\TeV}{\,{\rm TeV}}

\newcommand{\Tr}{\,{\rm Tr}}

\def\circa#1{\,\raise.3ex\hbox{$#1$\kern-.75em\lower1ex\hbox{$\sim$}}\,}

\newcommand{\beq}{\begin{equation}}
\newcommand{\eeq}{\end{equation}}

\newcommand{\bea}{\begin{eqnarray}}
\newcommand{\eea}{\end{eqnarray}}
\newcommand{\be}{\begin{equation}}
\newcommand{\ee}{\end{equation}}
\font\tenrsfs=rsfs10 at 12pt
\font\sevenrsfs=rsfs7 at 10 pt
\font\fiversfs=rsfs5
\newfam\rsfsfam
\textfont\rsfsfam=\tenrsfs
\scriptfont\rsfsfam=\sevenrsfs
\scriptscriptfont\rsfsfam=\fiversfs
\def\mathscr#1{{\fam\rsfsfam\relax#1}}

\def\Lag{\mathscr{L}}

\def\circa#1{\,\raise.3ex\hbox{$#1$\kern-.75em\lower1ex\hbox{$\sim$}}\,}
\makeatletter

\def\hhref#1{\href{http://arxiv.org/abs/#1}{arXiv:#1}} 

\def\hhref#1{\href{http://arxiv.org/abs/#1}{arXiv:#1}} 
 
\def\art{\@ifnextchar[{\eart}{\oart}}
\def\eart[#1]#2#3#4#5#6{{\rm #2}, {\em #3 \bf #4} {\rm (#6) #5} ({\em #1})}

\def\article{\@ifnextchar[{\earticle}{\oarticle}}
\def\oarticle#1#2#3#4#5#6{{\rm #1}, {\em ``#6''}, {\rm #2 #3 (#5) #4}}
\def\earticle[#1]#2#3#4#5#6#7{{\rm #2}, {\em ``#7''}, {\rm #3 #4 (#6) #5}  [\hhref{#1}]}
\def\hepart[#1]#2{{\rm #2, \em#1}}
\def\heparticle[#1]#2#3{#2, {\em ``#3''} [\hhref{#1}]}

\newcounter{alphaequation}[equation]
\def\thealphaequation{\theequation\hbox to
0.6em{\hfil\alph{alphaequation}\hfil}}
\def\eqnsystem#1{
\def\@eqnnum{{\rm (\thealphaequation)}}
\def\@@eqncr{\let\@tempa\relax \ifcase\@eqcnt \def\@tempa{& & &} \or
  \def\@tempa{& &}\or \def\@tempa{&}\fi\@tempa
  \if@eqnsw\@eqnnum\refstepcounter{alphaequation}\fi
\global\@eqnswtrue\global\@eqcnt=0\cr}
\refstepcounter{equation} \let\@currentlabel\theequation \def\@tempb{#1}
\ifx\@tempb\empty\else\label{#1}\fi
\refstepcounter{alphaequation}
\let\@currentlabel\thealphaequation
\global\@eqnswtrue\global\@eqcnt=0 \tabskip\@centering\let\\=\@eqncr
$$\halign to \displaywidth\bgroup \@eqnsel\hskip\@centering
$\displaystyle\tabskip\z@{##}$&\global\@eqcnt\@ne
\hskip2\arraycolsep\hfil${##}$\hfil& \global\@eqcnt\tw@\hskip2\arraycolsep
$\displaystyle\tabskip\z@{##}$\hfil
\tabskip\@centering&\llap{##}\tabskip\z@\cr}
\def\endeqnsystem{\@@eqncr\egroup$$\global\@ignoretrue} \makeatother

\newcommand{\SU}{\,{\rm SU}}
\newcommand{\SO}{\,{\rm SO}}
\newcommand{\U}{\,{\rm U}}
\newcommand{\Sp}{\,{\rm Sp}}

\definecolor{fiorentina}{rgb}{.5,0,.5}

\begin{document}

\centerline{IFUP-TH/2016\hfill CERN-PH-TH/2016-040\hfill ~~EFI-16-03 }

\bigskip
\bigskip

\begin{center}
{\LARGE \bf \color{rossos} 
Di-photon resonance and \\[3mm] Dark Matter as heavy pions}
\\[1cm]
\bigskip\bigskip

{\large\bf 
Michele Redi$^{a}$, Alessandro Strumia$^{b,c}$, \\[1mm]
Andrea Tesi$^d$, Elena Vigiani$^b$ 
}  
\\[5mm]

\bigskip

{\it $^a$ INFN, Sezione di Firenze, Via G. Sansone, 1, I-50019 Sesto Fiorentino, Italy}\\[1mm]
{\it $^b$ Dipartimento di Fisica dell'Universit{\`a} di Pisa and INFN, Italy}\\[1mm]
{\it $^c$ CERN, Theory Division, Geneva, Switzerland}\\[1mm]
{\it $^d$ Enrico Fermi Institute, University of Chicago, Chicago, IL 60637}\\[1mm]

\bigskip

\vspace{1cm}
{\large\bf\color{blus} Abstract}
\begin{quote}\large
We analyse confining gauge theories  where the 750 GeV di-photon resonance is a composite techni-pion
that undergoes anomalous decays into SM vectors. 
These scenarios naturally contain accidentally stable techni-pions Dark Matter candidates. 
The di-photon resonance can acquire a larger width by decaying into Dark Matter through the CP-violating  $\theta$-term 
of the new gauge theory reproducing the cosmological  Dark Matter density as a thermal relic.
\end{quote}

\thispagestyle{empty}
\end{center}

\setcounter{page}{1}
\setcounter{footnote}{0}

\newpage
\tableofcontents

\section{Introduction}
The simplest and most compelling explanation of the  $\gamma\gamma$ excess observed at $M_\X\approx 750 \GeV $~\cite{data} 
is provided by an $s$-channel scalar resonance $\X$ coupled to gluons and photons.

Theoretical analyses~\cite{1512.04933,toostrong} find that reproducing the experimentally favoured rate might need non-perturbative dynamics. 
Strongly interacting models elegantly predict resonances coupled to  $\gamma\gamma$ and gluons
(for example, they were mentioned in eq.~(95) of~\cite{strongDM},  before that the excess was found).
Loop-level decays into $\gamma\gamma$ and gluons give typically a small width. Taking into account that the ATLAS fit favours a resonance with a large width $\Gamma_\X\sim 0.06 M_\X$ (although with less than $0.5 \, \sigma$ improvement from the small width scenario),  extra decay channels could be needed. A suggestive  possibility is that the 750 GeV resonance has extra
decay channels into Dark Matter (DM) particles, given that these decays are relatively weakly constrained~\cite{1512.04933} and
that they allow to reproduce the observed cosmological DM abundance~\cite{DM750,1512.04933}.

We present simple explicit models where both the 750 GeV resonance $\X$ and DM are
Nambu-Goldstone bosons (NGB) of a new confining gauge theory, and where $\X$ can decay into DM pairs,
providing a relatively large width $\Gamma_\X$.

We will study confining gauge theories with fermions in a vectorial representation
of the SM, such that the new strong dynamics does not break the SM gauge group.
We assume that the Higgs is an elementary scalar particle.
As in QCD,  the lightest composite states are pion-like NGB arising from the spontaneous breaking of the 
accidental global symmetries of the new strong dynamics.\footnote{In the literature they are sometimes called `pions' or `techni-pions' or `hyper-pions': in order to avoid confusion and lengthy words
in the  text we will use TC$\pi$ for techni-pions, TCq for techni-quarks, $\Lambda_{\rm TC}$ for the dynamical scale, where techni-color (TC) refers to the new confining gauge interaction.}
The anomaly structure is entirely encoded in the Wess-Zumino-Witten term of the chiral Lagrangian,
giving rise to predictions for  the $\X$ decay rates into $\gamma\gamma$, $\gamma Z$, $ZZ$ and $gg$.

The interactions among TC$\pi$ are strongly constrained by the symmetries.
We will search for theories where  some of the TC$\pi$ are automatically long lived due to the accidental symmetries of the renomalizable Lagrangian
and provide DM candidates.\footnote{The strong dynamics  also produces accidentally stable techni-baryons that could be viable DM candidates~\cite{strongDM}. 
For techni-baryons made of light fermions the thermal production requires a dynamical scale in the 100 TeV range, incompatible with the di-photon excess. 
This conclusion could  be avoided with different production mechanisms or introducing  fermions heavier than the confinement scale. 
We will focus on techni-pions in this work.}
Two symmetries can be responsible for the stability of the DM techni-pions:
\begin{itemize}
\item{\em Species number}.  Models where TCq fill two copies $X_1$, $X_2$ of the same representation, 
give rise to neutral TC$\pi$ $\eta_\pm \sim X_1 \bar{X}_1 \pm X_2 \bar{X}_2$
which undergo anomalous decays to SM gauge bosons and to neutral TC$\pi$ $\Pi \sim X_1 \bar{X}_2$ stable
because of the accidental ${\rm U}(1)_1\otimes{\rm U}(1)_2$ symmetry thus providing automatic DM candidates.

\item{\em ${G}$-parity}. In models where TCq fill a representation $X$ plus its SM conjugate $\tilde{X}$,
one can impose a generalised $G$-parity symmetry that exchanges them.
As a consequence the lightest $G$-odd techni-meson  $\eta$ is a stable DM candidate~\cite{1005.0008,bai}.
This $G$-parity is not an accidental symmetry and can be broken by different mass terms.
Furthermore unbroken species number keeps stable the charged TC$\pi$ $\sim X \bar{\tilde X}$.

\end{itemize}


The paper is structured as follows. We start in section~\ref{2}  reviewing some general phenomenological aspects of the $\gamma\gamma$ excess.
In section~\ref{comp} we discuss the  structure of the theories and present the full list of models based on two SM species.
In section~\ref{3} we discuss general aspects of heavy pion DM phenomenology.
Models of composite DM  are discussed in section~\ref{4},
considering in section~\ref{sec:UNN} the case where DM stability results from species number,
and in section~\ref{G} models where DM is stable thanks to a $G$-parity.
In section~\ref{end} we present our conclusions. A technical appendix on the chiral Lagrangian in the 
presence of the $\theta$ angle follows.

\section{Phenomenology of the di-photon resonance}\label{2}


We will study theories where the 750 GeV resonance $\X$ is a composite pseudo-scalar
coupled to SM gauge bosons as described by the effective Lagrangian
\begin{equation}
\label{Lanomalies}
\Lag_{\rm WZW} \supset -\frac 1 {16\pi^2}\frac \X f\left[g_1^2\,c_B\, B_{\mu\nu}\tilde{B}^{\mu\nu}+g_2^2\,c_W\, W^a_{\mu\nu}\tilde{W}_a^{\mu\nu}+g_3^2\,c_G\, G^a_{\mu\nu}\tilde{G}_a^{\mu\nu}\right],
\end{equation}
where for a generic vector field $\tilde{V}_{\mu\nu}= \frac{1}{2} \epsilon_{\mu\nu\rho\sigma} V^{\rho\sigma}$. 
In models where $\X$ is a NGB $c_B,c_W,c_G$ are anomaly coefficients fixed by group theory, proportional to $N_{\rm TC}$ in $\SU(\NTC)$ gauge theories. 
In fact the full effect of anomalies  can be encoded in the Wess-Zumino-Witten term of the chiral Lagrangian that, up to the normalization only depends on the pattern of symmetry breaking.
The effective Lagrangian could also contain derivative couplings to SM fermion currents. 
This is for example the case in composite Higgs models with partial compositeness. In this work we  focus on UV complete theories based on gauge dynamics where such terms
do not appear at leading order so that it is sufficient for our analysis to focus on di-boson SM decay channels. 
In addition we consider the possibility that $\X$ can decay
in a extra  channel, $X$, focusing on the possibility that this is DM.
From the above Lagrangian,  the rate in  SM vector bosons is given by
\begin{equation}
\frac {\Gamma(\X\to VV)}{M_\X}= \kappa_V \frac {\alpha_V^2}{64\pi^3}c_V^2\frac {M_\X^2}{f^2}
\end{equation}
where $\kappa_V=1,8$ for photons and gluons and  $c_\gamma = c_B + c_W$. More explicitely  the rate into photons is
\begin{equation}
\frac {\Gamma_{\gamma\gamma}}{M_\X}= 3 \times 10^{-8}\, c_\gamma^2\frac {M_\X^2}{f^2} \,,
\label{eq:ratephotons}
\end{equation}
and the decay widths  into the other SM vectors are
\beq
\label{rates-anomalies}
\begin{array}{rclrcl}\displaystyle
\frac {\Gamma_{\gamma Z}}{\Gamma_{\gamma\gamma}}&\approx& \displaystyle
\frac {2(- c_W \cot \theta_W+ c_B \tan \theta_W)^2}{c_{\gamma}^2}\,,\qquad &\displaystyle
\frac {\Gamma_{Z Z}}{\Gamma_{\gamma\gamma}}&\approx&\displaystyle
\frac {(c_W \cot \theta^2_W+ c_B \tan \theta^2_W)^2}{c_{\gamma}^2},\\  \displaystyle
\frac {\Gamma_{WW}}{\Gamma_{\gamma\gamma}}&\approx&  
\displaystyle
2 \frac{c_W^2}{c_\gamma^2 \sin^4 \theta_W}\,, &\displaystyle
\frac {\Gamma_{gg}}{\Gamma_{\gamma\gamma}}&\approx& \displaystyle
\frac {8\alpha_3^2}{\alpha^2}\frac {c_{G}^2}{c_{\gamma}^2} \approx 1300 \, \frac {c_{G}^2}{c_{\gamma}^2}  \,.
\end{array}
\eeq
We assume in what follows a production cross-section,
\begin{equation}
\sigma(p p\to \X)_{13 \TeV} \times {\rm BR}(\X \to \gamma\gamma) \approx  5 \fb
\end{equation}
The experimental upper bounds on the other decay channels reads \cite{1512.04933}:
\begin{equation}
\frac {\Gamma_{\gamma Z}}{\Gamma_{\gamma\gamma}}< 5.6
\,,~~~~~ \frac {\Gamma_{Z Z}}{\Gamma_{\gamma\gamma}}< 12 
\,,~~~~~\frac {\Gamma_{WW}}{\Gamma_{\gamma\gamma}}< 40
\,,~~~~~\frac {\Gamma_{gg}}{\Gamma_{\gamma\gamma}}< 2500
\,,~~~~~\frac {\Gamma_{\rm DM}}{\Gamma_{\gamma\gamma}}< 800
\end{equation} 
implying the constraints on the anomaly coefficients
\begin{equation}
|c_G|< 1.4 |c_\gamma| \,,\qquad
 -0.3< \frac {c_W}{c_B} < 14   \,.
\end{equation}

\medskip

Assuming that the only 
relevant production channel is gluon fusion, as will always be the case in our models, the cross section is reproduced for~\cite{1512.04933}
\begin{equation}\label{eq:rate}
\frac {\Gamma_{\gamma\gamma}}{M_\X}\frac {\Gamma_{gg}}{M_\X} \approx  0.9  \times 10^{-9}\frac {\Gamma_\X}{\rm GeV} \,,
\end{equation}
that, combined with the latter equation of \eqref{rates-anomalies}, gives 
\begin{equation}
\frac{\Gamma_{gg}}{M_\X}= 1.1 \times 10^{-3}\, \frac {c_G}{c_\gamma} \sqrt{\frac {\Gamma_\X}{{\rm GeV}}}\,,~~~~~~~~~\frac{\Gamma_{\gamma\gamma}}{M_\X}= 0.84 \times  10^{-6} \,\frac {c_\gamma}{c_G}\sqrt{\frac {\Gamma_\X}{{\rm GeV}}} \,.
\label{eq:fit}
\end{equation}
From eq.~\eqref{eq:fit} and \eqref{eq:ratephotons} one can derive a relation between $f$ and the coefficients $c_\gamma$ and $c_G$:
\begin{equation}
\frac {M_\X} f \approx \frac {5.2} {\sqrt{c_\gamma c_G}} \left(\frac {\Gamma_\X}{{\rm GeV}}\right)^{\frac 1 4} \,,
\label{eq:moverf}
\end{equation}
implying that the $f$ is proportional to $\NTC$. 
Two cases are of special interest:
\begin{itemize}
\item{{\bf Small $\NTC$:} Maximises the strong sector effective coupling $g_{\rm TC}\sim 4\pi/\sqrt{\NTC}$ giving $M_\X/f \sim 10$.
The mass of the TC$\pi$  is around its maximal value $\Lambda_{\rm TC}\sim g_{\rm TC} f $. 
For example the $\eta'$ of QCD naturally falls into this category. Note that in this case states associated to
the new strong dynamics will be nearby. This is not necessarily a problem because  a large $g_{\rm TC}$ shields the strong dynamics effects.}

\item{{\bf Large $\NTC$:} Leads to a smaller strong coupling $g_{\rm TC}$, but the anomaly coefficients are enhanced by $\NTC$.
As a benchmark we can take $f\sim M_\X$, $\NTC\sim 10$.  The new strong dynamics now lies around 2-3 TeV but it is more strongly coupled to the SM.}
\end{itemize}
In what follows we will focus mostly on the first possibility. 
The second possibility implies a larger number of TCq, easily leading to Landau poles for SM couplings at low scales.

\begin{figure}[t]
\begin{center}
\includegraphics[width=.5\textwidth]{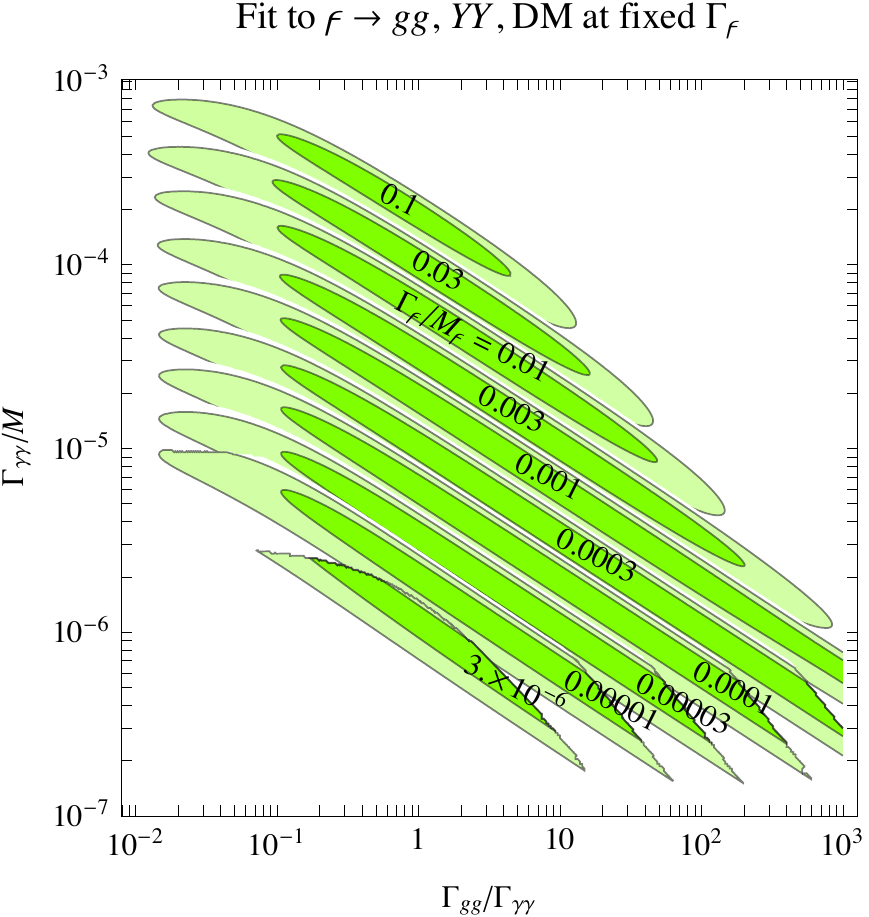}
\caption{\label{fig:GGrange}\em Values of $\Gamma_{gg}$ and $\Gamma_{\gamma\gamma}$ needed to reproduce the total width $\Gamma_\X$
indicated on the best-fit regions, assuming that $\X$ decays into $gg$, into hypercharge vectors, and into Dark Matter. }
\label{fig:maximumwidth}
\end{center}
\end{figure}

\subsection{Maximal $\X$ width}\label{GammaS}
In  absence of extra decay channels the di-photon signal requires \mbox{$\Gamma_{\gamma\gamma}/{M_\X}\approx 0.7 \times 10^{-6}$}, and the total width is dominated by $\Gamma_{gg}$.
The experimental bound on di-jets  implies
\begin{equation}
\Gamma_\X \approx \Gamma_{gg}< 2500 \, \Gamma_{\gamma\gamma}\approx 1.3 \,{\rm GeV}.
\end{equation}
A larger decay width needs new decay channels.
Let us assume that $\X$ decays to $\gamma\gamma$, to $gg$ and into a third channel $X$. 
We  have
\begin{equation}
\Gamma_{\gamma\gamma} \Gamma_{gg} = 5 \times 10^{-4}\, (\Gamma_{gg}+\Gamma_X)\,{\rm GeV}\,, \quad \Gamma_{gg} < 2500 \, \Gamma_{\gamma\gamma}\,, \quad \Gamma_X < k_X \Gamma_{\gamma\gamma}
\label{eq:maximumwidth}
\end{equation}
where the first equation demands that the total $pp\to \gamma\gamma $ rate is reproduced while the others are the experimental bounds 
on decays widths into $gg$ and $X$.
The most favourable situation is obtained when $X$ is DM: in such a case the experimental bound sets $k_X\approx 800$.  
The situation is summarized in fig.~\ref{fig:maximumwidth} that shows, as a function of
$\Gamma_{gg}/\Gamma_{\gamma\gamma}$,
the value of $\Gamma_{\gamma\gamma}$ needed to achieve different values of the total width
$\Gamma_\X/M_\X$: we see that a large width can be reproduced only if $\Gamma_{\gamma\gamma}$ is itself
large and $\Gamma_{gg}/\Gamma_{\gamma\gamma}$ is not too large.
These considerations are  encoded in the following equation,
obtained from eq.\eq{maximumwidth} by expressing $\Gamma_{gg}$ and $\Gamma_{\gamma\gamma}$ in terms of the model parameters $c_G$ and $c_\gamma$:
\begin{equation}
\Gamma_\X\, \circa{<}\,\Gamma_\X^{\rm max} =\left[ 0.25 \left(\frac{k_X}{800}\right)^2 \, \frac{c_\gamma^2}{c_G^2} + 0.67 \, \frac{c_G^2}{c_\gamma^2} + 0.83 \, \left(\frac{k_X}{800}\right) \right]{\GeV}.
\label{eq:widthbound}
\end{equation}
We see that decays into DM can give $\Gamma_\X\sim 45\GeV$ provided that
$c_\gamma \sim 15 \, c_G$.
As we will see, one can build models where $c_G$ is small.
However, substituting  eq.~(\ref{eq:moverf}) we obtain that the maximum width is roughly realised for
\begin{equation}
\frac {M_\X} f \approx \frac {3.7}{c_G}\sqrt{\frac {k_X}{800}}  \,.
\end{equation}
Given that the NGBs must be lighter than $M\circa{<} 10 f$, one finds $c_G\circa{>} 0.4$. 
The width into photons needed to generate the maximal width \eqref{eq:widthbound} is approximately given by
\begin{equation}
\frac {\Gamma_{\gamma\gamma}}{M_\X} \approx 0.7 \times 10^{-6}+ 4 \times 10^{-7} \left(\frac {k_X}{800}\right) \frac {c_{\gamma}^2}{c_G^2} \,.
\end{equation}

To summarise a width of 45 GeV would require in the most optimistic case $c_G\sim 0.5$ and $c_\gamma \sim 8$. 
While the first condition could be realised  we find that the second is extremely difficult to achieve in concrete models.

\begin{table}[t]
\begin{center}
\begin{tabular}{|ccc|c|}
\hline 
 $\SU(3)_c$ & $\SU(2)_L$ & $\U(1)_Y$ & name  \\
\hline \hline
 $1$ & $1$ & $0$ & $N$   \\  
 $1$ & $1$ & $1$ & $E$   \\
1& $2$ & $-1/2$ & $L$  \\
$1$ & $3$ & $0$  & $V$  \\
$\bar{3}$ & $1$ & $1/3$ & $D$  \\ 
$\bar{3}$ & $1$ & $-2/3$ & $U$   \\
$3$ & $2$ & $1/6$  & $Q$  \\
$1$ & $3$ & $1$  & $T$  \\
$6$ & $1$ & $-2/3$  & $S$  \\
\hline
\end{tabular}
\end{center}
\caption{\em \label{tab:irreps}  SM representations arising from  the smallest multiplets of the $\SU(5)$  unified group.
We assign standard names used throughout the paper.
}
\end{table}

\section{Confining theories for the di-photon resonance}\label{comp}
The above  phenomenological analysis applies in general to theories where the 750 GeV resonance is a NGB.  In particular couplings to SM gauge bosons
through anomalies depend only on the pattern of symmetry breaking up to an overall coefficient.
In what follows we will study UV realisations of this framework in terms of 4 dimensional gauge theories. 

We will focus on $\SU(\NTC)$ gauge dynamics with $\NF$ techni-flavours\footnote{Extensions to $\SO(\NTC)$ and $\Sp(\NTC)$ can be constructed along the same lines, see \cite{strongDM}. 
Singlets di-photon candidates have  identical properties to the ones discussed here so that any $\SU(\NTC)$ model can be extended to these gauge groups.}.  
The dynamics of this theory is well known from QCD and can be also understood in the large $\NTC$ limit: the gauge theory is asymptotically free (provided the usual bound on the number of techni-flavours is satisfied) and confines at a scale $\Lambda_{\rm TC}$. 
In order to avoid severe constraints (common to old techni-colour theories) we consider fermions that are in a vectorial representation of the SM and in the fundamental $\NTC$ of $\SU(\NTC)$~\cite{Kilic:2009mi,strongDM,Compositegammagamma}
\begin{equation}\label{eq:vectorial}
\Q= \sum_{i=1}^{N_{\rm S}} \Q_i ,\qquad \Q_i= (\NTC, R_i)\oplus (\bar{N}_{\rm TC},\bar R_i)  \,,
\end{equation}
where $R_i$ denotes a generic SM representation and $N_{S}$ is the number of species
with mass below the confinement scale. 
For a given TCq $\Q_i$, we denote as $\tilde \Q_i$ the  representation obtained
exchanging $R_i$ with $\bar R_i$: they are inequivalent if $R_i$ is complex.
For simplicity we consider $R_i$ representations that can be embedded in the simplest $\SU(5)$ representations listed and named in table \ref{tab:irreps}.

The choice in eq.~\eqref{eq:vectorial} ensures that the vacuum configuration of the confining sector does not break the SM $\SU(3)_c\otimes \SU(2)_L\otimes \U(1)_Y$ symmetries. 
Assuming  QCD-like dynamics the strong interactions confine  and spontaneously break the chiral global symmetry as $\SU(\NF)_L\otimes \SU(\NF)_R\to \SU(\NF)$ at the scale $f$ given by
\begin{equation}
\Lambda_{\rm TC} \sim \frac{4\pi}{\sqrt{\NTC}} f \,.
\end{equation}
The number of techni-flavour is given by
\begin{equation}
\NF= \sum_{i=1}^{N_{\rm S}} {\rm dim}(R_i) \,,
\end{equation}
where $N_{\rm S}$ is the number of SM species.
This produces NGBs,
the TC$\pi$, which are $\Q\bar \Q$ composite and thereby fill the representation
\begin{equation}\label{RRbar}
\left[\sum_{i=1}^{N_{\rm S}} R_i \right]\otimes \left[\sum_{j=1}^{N_{\rm S}}  \bar{R}_j \right] \,.
\end{equation}
We denote the singlets TC$\pi$ as $\eta$.
Given that each $R_i\otimes \bar R_i$ contains a singlet, any model contains at least $N_{\rm S}$ $\eta$ singlets.\footnote{Extra singlets exist if a  fermion representation appears with a multiplicity. These  singlets have no anomalies with SM gauge bosons and can be stable because of accidental symmetries.} Among them, the singlet associated with the generator proportional to the identity in techni-flavour space is anomalous under the $\SU(\NTC)$ gauge interactions.  Analogously to the  $\eta'$ in QCD, it acquires a large mass that can be estimated in a large-$\NTC$ expansion~\cite{veneziano} as
\begin{equation}
m_{\eta'}^2\sim \frac {\NF}{\NTC} \, \Lambda_{\rm TC}^2 \,,  
\label{eq:massetap}
\end{equation}
while the orthogonal combination $\eta$  acquires mass only from the mass terms of the TCq,
\mbox{$m_\eta^2 \sim m_\Q \, \Lambda_{\rm TC}$}, and can be much lighter. 

\medskip

The anomaly coefficients of the singlets $\eta$ with SM gauge bosons are given by
\begin{equation}
\label{c-anomalies}
c_B=2\NTC\,{\rm Tr} (T_\eta Y^2)\,,\quad c_W \, \delta^{ab}=2\NTC\,{\rm Tr}\, (T_\eta T^a T^b) \,,\quad 
c_G \, \delta^{AB}=2\NTC\, {\rm Tr} \,(T_\eta T^A T^B).
\end{equation}
Furthermore $c_\gamma =2\NTC\, \Tr( T_\eta Q^2)=c_B +c_W$. Here $T^a$ are the $\SU(2)_L$ generators, 
$T^A$ are the $\SU(3)_c$ generators, and $T_\eta$ is the chiral symmetry generator associated to the singlet $\eta$.

A remarkable feature of gauge theories is the existence of accidental symmetries. To each irreducible representation of fermions we can associate a conserved 
species number. This conserved quantum number is responsible for the accidental stability of TC$\pi$ made of different species. Discrete symmetries could also produce 
stable particles. In section \ref{4} we will construct explicit examples where stable TC$\pi$ are  identified with DM.

\begin{center}
\begin{sidewaystable}
\caption{\small \em Anomaly coefficients  for the $\eta$ and $\eta'$ singlets predicted by confinement models with two species compatible with experimental bounds. Models are selected requiring asymptotic freedom for $\SU(\NTC)$ and that the SM gauge couplings do not develop Landau poles below $\sim 10^{16} \GeV$ for $\NTC=3$. 
Models highlighted in green contain two acceptable di-photon candidates, the $\eta$ and the heavier $\eta'$,
while in blue (red) models only the $\eta'$ (the $\eta$) is an acceptable candidate,
while the other lighter (heavier) state is experimentally allowed. 
The prediction for $f/N_{\rm TC}$ is computed according to eq.~\eqref{eq:moverf}. 
Mixing between $\eta$ and $\eta'$ can modify the result.
}
\label{table:vectorlike}
\vspace{0.5cm}
\begin{small}
\begin{tabular}
{|c|c||c|c|c|c|c|c|c||c|c|c|c|c|c|c|}
\hline
$\Q$& $N_{\rm TF}$ & $\frac {c_B^\eta}{\NTC}$\,&  $\frac{c_W^\eta}{\NTC}$\,&$\frac{c_G^\eta}{\NTC}$\,  & $\frac{\Gamma^\eta_{\gamma Z}}{\Gamma^\eta_{\gamma\gamma}}$ & $\frac{\Gamma^\eta_{Z Z}}{\Gamma^\eta_{\gamma\gamma}}$ & $\frac{\Gamma^\eta_{GG}}{\Gamma^\eta_{\gamma\gamma}}$  & $\frac{f(\!\GeV)}{\NTC}$ 
& $\frac{c_B^{\eta'}}{\NTC}$\,& $\frac{c_W^{\eta'}}{\NTC}$\,& $\frac{c_G^{\eta'}}{\NTC} $\,  & $\frac{\Gamma^{\eta'}_{\gamma Z}}{\Gamma^{\eta'}_{\gamma\gamma}} $ & $\frac{\Gamma^{\eta'}_{Z Z}}{\Gamma^{\eta'}_{\gamma\gamma}}$ & $\frac{\Gamma^{\eta'}_{GG}}{\Gamma^{\eta'}_{\gamma\gamma}}$   & $\frac{f(\!\GeV)}{\NTC}$ 
\\ \hline \hline
\rowcolor[cmyk]{0,0.1,0,0} $D\oplus L$ & 5	&  ${\frac 1 6} \sqrt{\frac 5 3}$ &  ${\frac 1 2} \sqrt{\frac 3 5}$  & $-\frac 1 {\sqrt{15}}$ & 1.8 & 4.7 & 240  & 96 
&  ${\frac 1 3}  \sqrt{\frac 5 2}$ & $\frac 1 {\sqrt{10}}$ & $\frac 1 {\sqrt{10}}$ & 0.23 & 1.9 & 180   & $-$ 
\\
\rowcolor[cmyk]{0.1,0,0,0} $D\oplus U$ & 6		&  $\frac 1 {\sqrt 3}$ &  $0$  & $0$ & 0.57 & 0.082 & 0 & $-$   
& $\frac 5 {3 \sqrt{3}}$ & $0$ & $\frac 1 {\sqrt{3}}$ & 0.57 & 0.082 & 470  & 150  \\    
\rowcolor[cmyk]{0,0.1,0,0} $D\oplus E$	& 4	&  $\frac 4 3 \sqrt{\frac 2 3}$ &  $0$  & $-\frac 1 {2\sqrt{6}}$ & 0.57 & 0.082 & 46  & 170 
& $\frac {2 \sqrt{2}} 3$ & $0$ & $\frac 1 {2\sqrt{2}}$ & 0.57 & 0.082 & 180  & $-$
\\ 
\rowcolor[cmyk]{0.1,0,0,0}  $D\oplus Q$ & 9		& $-\frac 1 {6}$ &$\frac 1 {2}$ & $0$ & 17  & 22 & 0  & $-$
& $\frac 1 {3 \sqrt{2}}$ & $\frac 1 {\sqrt{2}}$ & $\frac 1 {\sqrt{2}}$ & 2.9 & 6.1 & 740  & 150 \\   
\rowcolor[cmyk]{0,0.1,0,0} $D\oplus T$ & 6	      &   $\frac 8 {3 \sqrt{3}}$ &$\frac 2 {\sqrt{3}}$ & $-\frac 1 {2 \sqrt{3}}$ & 0.43  & 2.4 & 15   & 430
&  $\frac {10} {3 \sqrt{3}}$ & $\frac 2 {\sqrt{3}}$ & $\frac 1 {2\sqrt{3}}$ & 0.23 & 1.9 & 12  & $-$ 
\\
\rowcolor[cmyk]{0.1,0,0,0} $L\oplus U$	& 5	& $\frac 7 {6 \sqrt{15}}$ & $-{\frac 1 2} \sqrt{\frac 3 5}$ & $\frac 1 {\sqrt{15}}$ & 200 & 180 & 12000 & $-$ 
& $\frac {11} {3\sqrt{10}}$ & $\frac 1 {\sqrt{10}}$ & $\frac 1 {\sqrt{10}}$ & 0.0027 & 0.83 & 60  & 230  \\    
\rowcolor[cmyk]{0.1,0,0.25,0} $L\oplus Q$	& 8	& $-\frac 2 {3\sqrt{3}}$ &$0$ & $\frac 1 {2  \sqrt{3}}$ & 0.57  & 0.082 & 740   & 61
& $\frac 1 3$ & $1$ & $\frac 1 2$ & 2.9 & 6.1 & 180  & 210 
\\   
\rowcolor[cmyk]{0.1,0,0,0} $L\oplus S$ & 8 &  $\frac 7 {12\sqrt{3}}$ & $- \frac {\sqrt{3}} 4$ & $\frac 5 {4\sqrt{3}}$ & 200  & 180 & 74000 & $-$ 
&  $\frac {19} {12}$ & $\frac 1 4$ & $\frac 5 4$ & 0.095 & 0.47 & 610   & 290  
\\   
\rowcolor[cmyk]{0.1,0,0.25,0} $U\oplus E$ & 4 & $\frac 5 {3\sqrt{6}}$ & 0 & $-\frac 1 {2\sqrt{6}}$ & 0.57  & 0.082 & 120    & 110
& $\frac 7 {3 \sqrt{2}}$ & 0 & $\frac 1 {2\sqrt{2}}$ & 0.57 & 0.082 & 60  & 260 
\\   
\rowcolor[cmyk]{0.1,0,0,0} $U\oplus Q$ & 9	& $-\frac 5 6$ & $\frac 1 2$ & $0$ & 32  & 17 & 0  & $-$ 
& $\frac 1 {\sqrt{2}}$ & $\frac 1 {\sqrt{2}}$ & $\frac 1 {\sqrt{2}}$ & 0.79 & 3.0 & 330  & 220  \\ 
\rowcolor[cmyk]{0.1,0,0,0} $U\oplus V$ & 6	& $-\frac 4 {3\sqrt{3}}$ & $\frac 2 {\sqrt{3}}$ & $-\frac 1 {2\sqrt{3}}$ & 83  & 82 & 740 & $-$  
& $\frac 4 {3\sqrt{3}}$ & $\frac 2 {\sqrt{3}}$ & $\frac 1 {2\sqrt{3}}$ & 1.5 & 4.1 & 29  & 310 \\  
\rowcolor[cmyk]{0,0.1,0,0} $U\oplus N$ & 4	& $-{\frac 2 3} \sqrt{\frac 2 3}$ & 0 & $-\frac 1 {2\sqrt{6}}$ & 0.57  & 0.082 & 180  & 87
& $\frac {2 \sqrt{2}} 3$ & 0 & $\frac 1 {2\sqrt{2}}$ & 0.57 & 0.082 & 180  & $-$ 
\\
\rowcolor[cmyk]{0,0.1,0,0} $E\oplus Q$ & 7	& $-{\frac 5 6} \sqrt{\frac 7 3}$ & ${\frac 1 2} \sqrt{\frac 3 7}$ & $\frac 1 {\sqrt{21}}$ & 3.6  & 0.52 & 70   & 150
& ${\frac 1 3} {\sqrt{\frac 7 2}}$ & $\frac 3 {\sqrt{14}}$ & $\sqrt{\frac 2 7}$ & 1.2 & 3.7 & 180  & $-$ 
\\
\rowcolor[cmyk]{0.1,0,0.25,0} $E\oplus S$ & 7	& $-\frac {10} {3 \sqrt{21}}$ & $0$ & $\frac 5 {2\sqrt{21}}$ & 0.57  & 0.082 & 740 & 120
& ${\frac {11} 3} \sqrt{\frac 2 7}$ & $0$ & $5 \sqrt{14}$ & 0.57 & 0.082 & 610  & 310
\\   
\rowcolor[cmyk]{0.1,0,0,0} $S\oplus V$ & 9 &  $-\frac 8 9$ & $\frac 4 3$ & $- \frac 5 6$ & 83 & 82 & 4600 & $-$ 
& $\frac {8 \sqrt{2}} 9$ & $\frac {2\sqrt{2}} 3$ & $\frac 5 {3\sqrt{2}}$ & 0.43 & 2.4 & 380  & 350 \\   
\rowcolor[cmyk]{0.1,0,0.25,0}  $S\oplus N$	& 7	&  $-\frac 8 {3\sqrt{21}}$ & $0$ & $-\frac 5 {2\sqrt{21}}$ & 0.57 & 0.082 & 1200  & 93
& ${\frac 8 3} \sqrt{\frac 2 7}$ & $0$ & $\frac 5 {\sqrt{14}}$ & 0.57 & 0.082 & 1200 & 230  \\ 
\rowcolor[cmyk]{0.1,0,0.25,0}  $N\oplus Q$	& 7	&  $\frac{1}{6\sqrt{21}}$ & $\frac{1}{2} \sqrt{\frac 3 7}$ & $\frac 1 {\sqrt{21}}$ & 4.9 & 8.5 & 470  & 58 
& $\frac 1 {3 \sqrt{14}}$ & $\frac 3 {\sqrt{14}}$ & $\sqrt{\frac 2 7}$ & 4.9 & 8.5 & 470 & 140  
\\ \hline
\end{tabular}
\end{small}
\end{sidewaystable}
\end{center}

\subsection{Models with two species}


In table \ref{table:vectorlike} we give a full list of models with two TCq, that is $\Q=X_1+X_2$,
that can be embedded into unified representations and remain perturbative up to the unification scale
These models provide 2 di-photon candidates for $\X$, the $\eta$ and $\eta'$. 

Asymptotic  freedom of the $\SU(\NTC)$ gauge theory and absence of Landau poles for SM couplings 
below the unification scale allow only a finite list of possibilities.
These models do not contain DM candidates, so that the width is dominated by $\Gamma_{gg}$. They can be extended to contain DM
candidates by adding fermions that are singlets under the SM, see section \ref{4}.

Notice that the anomaly computation is reliable for $m_\pi \ll \Lambda_{\rm TC}$: for $f\sim 100 $ GeV, $\X$ is close to the cut-off of the 
effective Lagrangian and higher dimensional operators could give important contributions. In QCD the $\eta$ and $\eta'$
decay widths are predicted with $30\%$ precision from the anomaly computation. We therefore consider the values 
in table \ref{table:vectorlike} as an estimate with an error of similar size. 

The singlets $\eta$ and $\eta'$ in general mix.  Their mixing can be estimated from the chiral Lagrangian as,
\begin{equation}
\theta_p \sim \frac{\Lambda_{\rm TC} (m_{\Q_1}-m_{\Q_2})}{m_{\eta'}^2 - m_{\eta}^2}   \,.\end{equation}
As a consequence their anomaly coefficients correspondingly mix.
For $N_{\rm TC}\sim 3$ the mass of $\eta$ and $\eta'$ are comparable so that the  mixing can be significant. 
For example in QCD the mixing angle between $\eta(550)$ and $\eta'(958)$ is estimated around $-15^\circ$ \cite{PDG}  in rough agreement with the formula above.

In the models of table \ref{table:vectorlike} where the di-photon candidate is the $\eta'$ singlet, the value of $f$ suggests a mass scale for the $\eta'$ above $750 \GeV$. We note however that  the estimate of the mass and coupling to photons of the $\eta'$ is particularly uncertain away from the QCD case with 3 colours and 3 flavours.
In any case, consistency with the di-photon signal can be recovered thanks to a mixing between $\eta$ and $\eta'$. A sizable mixing is indeed common since in order to avoid the experimental constraints on extra coloured particles, TCq masses should not be much smaller than the confinement scale, so that the $\eta$ and $\eta'$ have comparable mass allowing them to significantly mix.

The $\eta/\eta'$ mixing can give an accidentally small $c_G$ for the 750 GeV resonance. In models with other decay channels such as DM
this could allow to increase the total width as discussed in section~\ref{GammaS}.

\subsection{Other Resonances}
The phenomenology of confinement models is  rich and has been discussed for example in \cite{Kilic:2009mi,strongDM}.
Given the fermion content, quantum numbers of the resonances are predicted. In a QCD-like theory the lowest lying states are expected to be techni-pions and spin-1 resonances (TC$\rho$) at a higher mass $\Lambda_{\rm TC}$. 

Before discussing the techni-pions we consider the TC$\rho$. Differently from the TC$\pi$, the interactions and the mass scale of the TC$\rho$ are less calculable. 
There is however a universal feature: coupling with the SM fermions arises through the mixing with SM gauge bosons, such that the resulting strength scales as
\begin{equation}
g \sim \frac {g_{\rm SM}^2}{g_{\rm TC}} \,.
\end{equation}
The coupling $g$ of the techni-resonances to the SM fields is suppressed by the large value of $g_{\rm TC}$, especially for $\NTC=3$.
Thanks to this generic fact, models with $\Lambda_{\rm TC}\sim $ TeV are experimentally allowed. 

Let us turn to techni-pions. A colour anomaly requires the existence of fermion constituents with colour so that all models 
predict coloured scalars with mass around the di-photon resonance. Moreover in models with more than one specie there are 
extra singlets that also couple to gluons and photons through anomalies and could be singly produced at the LHC.

\begin{figure}[t]
\begin{center}
\includegraphics[width=0.45\textwidth]{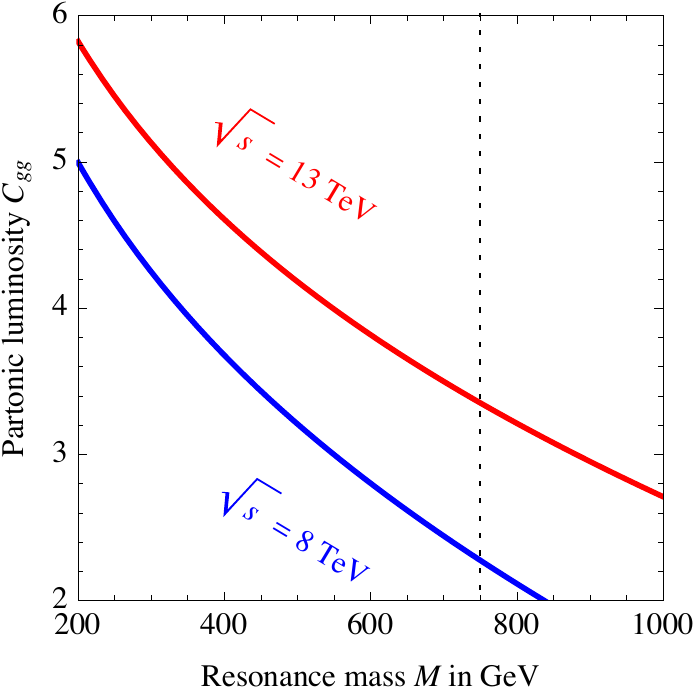}\qquad
\includegraphics[width=0.45\textwidth]{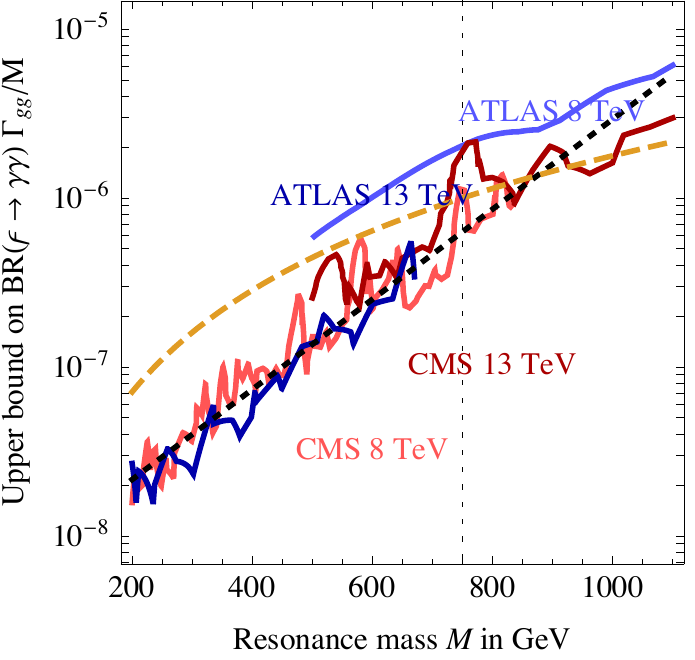}
\caption{\label{fig:Cggmass}\em {\bf Left:} partonic luminosities for $gg$-initiated $pp\to \X$ scattering at different $\X$ masses.
{\bf Right}: experimental bounds on $pp\to \X\to\gamma\gamma$ \cite{data,Aad:2015mna,Khachatryan:2015qba}.}
\end{center}
\end{figure}

\subsubsection*{Extra techni-$\eta$ singlets}
\label{extraeta}
The candidate for the di-photon resonance with mass $M_\X\approx 750\GeV$ is accompanied by $N_{\rm S}-1$  extra singlets,
lighter or heavier.  Their couplings to the SM vectors are again
described by a Lagrangian of the same form as eq.~(\ref{Lanomalies}).  Assuming that they couple to gluons, their production cross section is
\begin{equation}\label{eq:extrasingletXS}
\sigma(p p\to \eta ) =  C_{gg}(m_\eta) \,\frac{\Gamma_{gg}}{m_\eta\,s}\,
\end{equation}
where $s$ is the collider energy and $C_{gg}(m_\eta)$ are the dimensionless partonic luminosity for the single production  of a resonance with mass $M=m_\eta$ from $gg$ partons in $pp$ collisions,
\begin{equation}
C_{gg}(M) = \frac{\pi^2 }{8} \int_{M^2/s}^1 \frac{dx}{x} g(x) g\left(\frac{M^2}{sx}\right).
\end{equation}
The numerical value of $C_{gg}$ as a function of the mass is shown in fig.~\ref{fig:Cggmass}a.
The experimental bounds on $pp\to\gamma\gamma$ are given in fig.~\ref{fig:Cggmass}b,
and can be roughly approximated as (dotted curve in  fig.~\ref{fig:Cggmass}b)
\beq \frac{\Gamma_{gg}}{M}\times\hbox{BR}( \eta \to\gamma\gamma)\circa{<}10^{-8.2 + 2M/M_\X}.\eeq
The left-handed side can be approximated as $\Gamma_{\gamma\gamma}/M$, in models where $\Gamma_{\gamma\gamma}\ll  \Gamma_{gg}\approx\Gamma_\eta$.
For given anomaly coefficients $c_B,c_W,c_G$, the branching ratios do not depend on the mass
$M$ and the widths scale as $\Gamma_{gg}\propto M^3$
(dashed curve in  fig.~\ref{fig:Cggmass}b)
as long as $M\gg M_Z$.
This means that, in the simple relevant limit where $\Gamma_{gg}\gg \Gamma_{\gamma\gamma}$,
the experimental bounds on $c_{\gamma}$  are about a factor of 2.5
stronger at $M=\frac12 M_\X$ with respect to $M=M_\X$.

The presence of one or two di-photon candidates and the compatibility of the $pp \to \gamma\gamma$ bound distinguishes the models of table 2 in three categories, denoted with different colours. 
The models highlighted in green contain 2 di-photon candidates, which are both acceptable candidates for the 750 GeV resonance.
The models in blue contain only one acceptable di-photon candidate (the $\eta'$) and a lighter singlet $\eta$ that is compatible with the experimental bound of fig.~\ref{fig:Cggmass}b. In some models the $\eta$ does not couple to gluons, so that its production is strongly suppressed. 
The models in red contain only one acceptable di-photon candidate (the $\eta$), and a heavier singlet $\eta'$.

\subsubsection*{Extra coloured techni-pions}\label{8}
Techni-pions in a real representation of the SM can decay into SM vector.
We consider the single production of a coloured $\chi=(8,1)_0$ that mainly decays to $jj$, with
cross-section given by
\begin{equation}
\sigma(p p\to \chi \to jj) = \frac {8 \, C_{gg}(m_\chi)}  {m_\chi s}  \, \Gamma(\chi \to jj) \,  \mathrm{BR}(\chi \to jj)
\end{equation}
where the quantities are defined as in eq.~\eqref{eq:extrasingletXS}. The interaction term 
\begin{equation}
-\frac{g_3^2}{32\pi}\, \NTC \, d^{abc}\frac{\chi^a}{f}\, G_{\mu\nu}^b \tilde{G}^{c,\mu\nu},\,\quad  d_{abc}=2 \mathrm{Tr}[T_a\{T_b,T_c\}]
\end{equation}
gives the decay width
\begin{equation}
\frac{\Gamma(\chi \to jj)}{m_\chi}= C_8\, \frac{\alpha_3^2}{2048\pi^3}\, \NTC^2\, \frac{m_\chi^2}{f^2}, \quad \quad C_8 = \sum_{abc} d_{abc}^2= \frac{40}{3}.
\end{equation}
Di-jet searches at $\sqrt{s}=8 \TeV$~\cite{Aad:2014aqa,CMS:2015neg} imply $f/N_{\rm TC}\circa{>}70\GeV$ for $\chi$ masses between
0.5 and $1.5 \TeV$. We therefore consider as a safe bound 1 TeV for the mass of the colour octet, since many models will require a value of $f/\NTC$ similar to the above in order to match the diphoton rate.
If composed of charged constituents,  $\chi$ also decays to $\gamma j$ and $Z j$, with branching ratios suppressed by $\sim \alpha/\alpha_3$:
these decay modes lead to weaker bounds.

\medskip


Complex TC$\pi$ are mainly produced via pair production. Limits on pair produced (8,1) and (8,3) TC$\pi$ are much weaker, although they are fairly model independent since the production is  determined by SM gauge interactions. A rough bound on a pair-produced colour octets decaying to pairs of $jj$ is $\approx450\ \mathrm{GeV}$~\cite{Khachatryan:2014lpa} (after matching to the production rate for colour octets). This bound is weaker than the one from single production, although it can be the dominant one for models with a large $f/\NTC$.   
TC$\pi$ charged only under the electro-weak group have smaller production cross section at the LHC.

\medskip

The experimental limits on coloured techni-pions, especially those from di-jet searches, potentially constrain some models of table  \ref{table:vectorlike}, however the actual bounds on a concrete model depends on the details of the mass spectrum. For a detailed study of the phenomenology of a given model, see \cite{nomura2} where the model $\Q=D\oplus L$ is considered.

\subsection{Effective Lagrangian}
\label{sec:CPviolation}
The interactions of the TC$\pi$ can be studied using chiral Lagrangian techniques, reviewed in the appendix, to which we refer for all the details. We include in our description the $\eta'$
that provides a di-photon candidate in most models. 
Of particular relevance to the following discussion will be the hidden sector $\theta_{\rm TC}$ angle (see also \cite{Draper:2016fsr}).
The strong dynamics violates CP if its action includes the topological term
\begin{equation}
\frac {\theta_{\rm TC}}{16\pi^2} \int d^4 x\, {\rm Tr} \,[{\mathcal G}_{\mu\nu}\tilde {\mathcal G}^{\mu\nu}] \,,
\end{equation}
$\theta_{\rm TC}$ is physical if the masses of the TCq are different from zero.  
We assume in what follows that the QCD strong CP problem is solved by axions in the usual way and that no axion mechanism exists for $\theta_{\rm TC}$\footnote{As noted in \cite{nomura2} the QCD axion does not eliminate contributions to the Weinberg operator that also contributes to the neutron EDMs. Using  NDA estimate one finds that this contribution is compatible with present bounds for a large region of parameters.}. 

On the other hand $\theta_{\rm TC}$ has important effects on the spectrum and dynamics of the composite states.
The main physical effects of $\theta_{\rm TC}$ is to induce electric dipoles for the techni-baryons \cite{strongDM} 
and CP-violating interactions for techni-pions \cite{derafael}. The latter is important in the present context as it allows the decay 
of the 750 GeV di-photon candidate  $\eta$ into lighter TC$\pi$ pairs.  For the present work it will be sufficient the following effective Lagrangian \cite{derafael}
\bea
\label{eq:CPviolation}
\Lag_{\rm eff} &=& \frac{f^2}{4} \Bigl\{ \Tr \left[ (D_\mu \tilde{U}) (D^\mu \tilde{U})^\dag \right] + \Tr \left[2 B_0 {\mathcal M}_\Q (\tilde{U} + \tilde{U}^\dag)\right]  - \frac{a}{\NTC} \Big[\frac{i}{2} \, {\rm log} \Big(\frac{{\rm det}\, \tilde{U}}{{\rm det} \, \tilde{U}^\dag}\Big) - \theta_{\rm TC} \Big]^2 \Big\} \notag \\
&&+ \Lag_{\rm WZW} \,,
\eea
written in terms of the field $\tilde{U}(x)\equiv \langle \tilde{U}\rangle U(x)$, where $U(x)={\rm exp}(-i 2 \Pi(x)/f)$,  $\Pi(x)$ is the TC$\pi$ matrix  including the $\eta'$ and $\tilde{U}$ is a diagonal unitary matrix. The matrix ${\mathcal M}_\Q={\rm Diag}[m_i]$ includes all the TCq masses, $B_0$ is a non perturbative constant of order ${\mathcal O}(\Lambda_{\rm TC})$ and $a$ is related to the $\eta'$ mass as $m_{\eta'}^2\approx \NF a /\NTC+ {\cal O}({\mathcal M_\Q})$.
For $\theta_{\rm TC}\neq 0$ the vacuum is at $\langle \tilde{U}\rangle\neq \One$ and the minimization of the potential leads to the Dashen's equations, see eq. \eqref{dashen} in the appendix. 
Expanding around the vacuum one finds cubic vertices for the techni-pions
\begin{equation}
\Lag_{\rm cubic} = \frac{2a}{3 \NTC f} \, \bar\theta_{\rm TC} \, \Tr[ \Pi^3 ] 
\label{eq:trilinear}
\end{equation}
where $\bar{\theta}_{\rm TC}$  measures the violation of CP and is related to the TCq masses 
and the $\theta_{\rm TC}$-angle by the Dashen equations. For small fermion masses the approximate relation
\begin{equation}
\frac{a}{\NTC} \, \bar\theta_{\rm TC} \sim m_{\rm min}  \, \Lambda_{\rm TC} \, \theta_{\rm TC}
\end{equation}
holds for small $\theta_{\rm TC}$. Accurate formulas can be found in  the appendix.

\medskip

Techni-pions have also multipole couplings to SM gauge bosons that are of phenomenological relevance.
This is particularly important for neutral techni-pions that do not couple to SM fields to leading order.
Such couplings explicitly break the global symmetries so they have to be proportional to the mass
parameters of the fundamental Lagrangian. The strong dynamics generates operators such as \cite{SILH}
\begin{equation}
\frac {g_3^2\,\NTC}{16\pi^2} \frac {1} {\Lambda_{\rm TC}} \, {\rm Tr}[{\mathcal M}_\Q \tilde{U}+{\mathcal M}_\Q \tilde{U}^\dagger] \,G_{\mu\nu}^a G^{a,\mu\nu}.
\end{equation}
Analogous couplings to electro-weak gauge bosons are also generated.
Expanding this term one finds CP preserving interaction $({\rm TC}\pi)^2 \, G^2$ and well as CP violating terms ${\rm TC}\pi \, G^2$ further suppressed by
$\bar{\theta}_{\rm TC}$.  The first ones, also known as Cromo-Rayleigh interactions, will play an important role in the DM phenomenology discussed in the next section~\cite{Bai:2015swa}. 
They also allow double production of the di-photon candidate through gluon fusion. From the above equation the coupling can be estimated as
\begin{equation}
\frac {g_3^2\,\NTC}{16\pi^2}\frac {M_\X^2}{\Lambda_{\rm TC}^2} \frac {\eta^2}{f^2}  G_{\mu\nu}^a G^{a\mu\nu}.
\label{eq:rayleigh}
\end{equation}
CP-violating effects in $\eta$ decays to SM gauge bosons are further suppressed, see also \cite{Draper:2016fsr}.

\section{Phenomenology of techni-pion Dark Matter}\label{3}
Gauge theories automatically deliver  particles stable thanks to accidental symmetries.
In particular in models with several SM representations, TC$\pi$ made of different species are stable at the renormalisable level.
Alternatively TC$\pi$ could be stable imposing appropriate discrete symmetries.
It is tempting to identify such particles with DM.

DM as a composite scalar TC$\pi$ can be charged or neutral under the SM gauge group.
In the former case SM gauge interactions contribute to the DM annihilation cross section
as in minimal DM models \cite{MDM,strongDM}, such that, for DM masses below a TeV,
the thermal relic DM abundance is smaller than the observed cosmological DM abundance.
The only possible exception is $N\ge 2$ copies of scalar doublets.

We focus in what follows on neutral DM candidates, that we will call $\Pi$.
From a phenomenological point of view, their most relevant interactions are with
gluons ~\cite{Bai:2015swa} and with the di-photon resonance $\eta$ as described in the previous section. 
The leading terms relevant for the DM interactions are\footnote{When $\Pi$ is not the lightest TC$\pi$ or others almost degenerate TC$\pi$ exist, co-annihilations with TC$\pi$ in thermal equilibrium with the SM can provide a more efficient mechanism for thermal production, making the previous interactions subleading (although they still play a role in detection experiments). The dominant process is TC$\pi$ scattering from 4-point interactions arising from the first and second term in eq.~\eqref{eq:CPviolation}.} 
\begin{equation}\label{eq:DM-interactions}
\begin{split}
&\Lag_{\rm DM} = C_{\eta\Pi\Pi}  \frac{\eta\Pi^2}{2} - \frac {g_3^2} {16\pi^2} \, c_G\,\frac {\eta}{f}\, G_{\mu\nu}^a\tilde{G}^{a,\mu\nu}+ \frac{g_3^2}{16\pi^2  }C_{\Pi\Pi gg} \frac {\Pi^2}{f^2}\,G_{\mu\nu}^a G^{a,\mu\nu}\,, 
\end{split}
\end{equation}
%

The NGB nature of the particles implies restrictions on the coefficients of the effective operators. Since the above operators break the NGB shift symmetry their coefficient must be proportional to the explicit breaking effects. While for the $\eta$ the coefficient $c_G$ is due to the strong interactions, for stable singlets like $\Pi$ the only source of explicit breaking is given by the fermion masses so that the coefficients above must be proportional to the TC$\pi$ mass.  Moreover, $C_{\eta\Pi\Pi}$ breaks  both the shift symmetry and CP so that it is proportional to the TC$\pi$ mass and to $\theta_{\rm TC}$.
From eq.~(\ref{eq:trilinear}) and (\ref{eq:rayleigh}) one finds the  estimates,
\begin{equation}\label{eq:DM-interactions-NDA}
C_{\Pi\Pi gg} \sim \NTC  \frac {m_\Pi^2}{\Lambda_{\rm TC}^2}\,,~~~~~~~~~~ C_{\eta\Pi\Pi}\sim \frac {m_\Pi^2}{f} \theta_{\rm TC} \,. 
\end{equation}
As expected the coefficients go to zero for $m_\Pi\to 0$ as in this limit $\Pi$ becomes an exact NGB. 
Our estimate differs from \cite{bai} where the coefficient was assumed to be constant. In models with lighter coloured NGB, $\Lambda_{\rm TC}$ should be replaced by the mass of these objects as a perturbative computation shows. Coloured resonances should however be heavier than about 1 TeV. The coefficient $c_G$ and $C_{\eta\Pi\Pi}$ can be extracted from the leading terms of the chiral Lagrangian, while the coefficients of the Rayleigh interaction  can only be estimated. Therefore, when the dominant interactions between DM and the SM are induced by the cubic CP-violating couplings, this setup is calculable.

\subsection{Thermal relic abundance}
Assuming that the  interactions in eq.~\eqref{eq:DM-interactions} dominate, the thermal relic abundance of DM can be derived in the standard way.
From the $s$-wave annihilation cross section of a real scalar DM we obtain:
\begin{equation}\label{eq:sigmav}
\langle \sigma v \rangle = \frac{\alpha_3^2}{\pi^3}\frac{m_\Pi^2}{f^4} \left[ 4 C_{\Pi\Pi gg}^2 +  \frac{C_{\eta\Pi\Pi}^2 c_G^2 f^2}{(M_\X^2-4m_\Pi^2)^2 + M_\X^2 \Gamma_\X^2}
 \right] + {\cal O}(v^2).
\end{equation}
If the first non-resonant contribution dominates,  the observed relic abundance is reproduced for
\begin{equation}
m_\Pi \sim 600 \GeV \left(\frac{f}{400\ \mathrm{GeV}}\right )^2 \left(\frac{0.3}{C_{\Pi\Pi gg}}\right)  \,.    
\end{equation}
The second contribution is generically expected to be comparable and it can be resonantly enhanced if 
$m_\Pi \approx \frac12 M_\X$.

The situation is illustrated in figure \ref{fig:UNN}, where along the solid blue curves the relic abundance is mainly reproduced due to CP-violating effects from $\theta_{\rm TC}$, for the models of sections \ref{sec:UNN} and \ref{sec:Baimodel}.

\bigskip

In some models (see section \ref{sec:UNN}) an extra singlet $\eta_*$ is  {lighter than DM}, or almost degenerate 
with it, and decays into SM vectors through anomalies. This extra light state changes the thermal relic abundance with respect to our discussion above. 
Interactions between DM and $\eta_*$ arise from the non-linearities of the kinetic term and mass terms and have the generic form
\begin{equation}
\Lag \sim \frac 1 {f^2} \eta_*^2 (\partial \Pi)^2 + \frac {m^2}{f^2}  \eta_*^2  \Pi^2.
\end{equation}
The DM annihilation cross section receives an extra contribution from DM DM $\to \eta_* \eta_*$ scatterings,
which can be estimated as
$ \sigma v  \approx \sfrac{m_\Pi^2}{64\pi f^4}$.  When this dominates, the desired thermal relic abundance is reproduced
for a DM mass $m_\Pi \sim 50 \GeV (f/300\GeV)^2$.
Each model predicts a specific form for these interactions: 
the model  in section \ref{sec:UNN} reproduces the thermal relic abundance along the
dashed blue curve in the right panel of figure \ref{fig:UNN}.
Such interactions will be (somewhat improperly) named co-annihilations, given that $\eta_*$ is part of the DM sector,
and in some limits $\eta_*$ itself becomes a stable DM particle.

\subsection{Direct Detection}
Integrating out the di-photon $\eta$, we obtain from eq.~\eqref{eq:DM-interactions} 
the effective interactions relevant for low-energy direct DM detection:
\be
\label{eq:DM-effective}
\Lag_{\rm eff} =
\frac{g_3^2}{16\pi^2} \frac{\Pi^2_*}{f^2}\bigg[
C_{\Pi\Pi gg}\ G_{\mu\nu}^a G^{a,\mu\nu}-\frac{C_{\eta\Pi\Pi} c_G\, f}{2 M_\X^2}\ G_{\mu\nu}^a \tilde G^{a,\mu\nu} \bigg] \,.
\ee
The first CP-conserving operator contributes to the spin-independent cross section as~\cite{Bai:2015swa}
\begin{equation}
\sigma_{\mathrm{SI}}= \frac{9 f^2_g}{4\pi}C_{\Pi\Pi g g}^2\frac{m_N^4}{(m_N+m_\Pi)^2 f^4}
\end{equation}
where $f_g = 2 (1 - f_u - f_d - f_s)/27\approx 0.064$
parameterizes the nucleon matrix element \cite{1110.3797, 1209.2870} and $m_N$ is the nucleon mass.
Numerically we get
\begin{equation}
\sigma_{\mathrm{SI}}= 0.16 \times 10^{-46} \mathrm{cm}^2\ 
\left(\frac{f_g}{0.064}\right)^2 
\left(\frac{C_{\Pi\Pi g g}}{0.1}\right)^2 
\left(\frac{300\ \mathrm{GeV}}{m_\Pi}\right)^2 
\left(\frac{300\ \mathrm{GeV}}{f}\right)^4 \,, 
\end{equation} 
which is in the interesting ballpark for future experiments.

\medskip

The other CP-violating operator induces a spin dependent coupling to the nucleons, further suppressed by the 
small exchanged momentum $\delta\vec{q}$:
\begin{equation}
\frac{d\sigma_{\mathrm{SD}}}{d\cos\theta}= \frac{\eta_N^2}{2\pi} \left(\frac{C_{\eta\Pi\Pi} c_G\,f}{ M_\X^2}\right)^2 \frac{m_N^2 |\delta\vec{q}|^2}{(m_N+m_\Pi)^2 f^4},
\end{equation}
where $\eta_N= (0.41,-0.0021)$ for $N=(p, n)$ \cite{Bai:2015swa}. For typical values of the parameters this cross-section is $\sigma_{\rm SD}\approx 10^{-47}\ \mathrm{cm}^2$,
 well below the current and future sensitivity.
 Similarly, DM indirect detection  is not significantly constrained, unless DM annihilations have a significant branching ratio into $\gamma\gamma$ lines~\cite{DM750}.

\subsection{Collider constraints}
The operators in eq.~\eqref{eq:DM-effective} can be also constrained by searches at the LHC. 
Assuming the validity of the effective operator description, namely that the mediator is sufficiently heavy,
the bounds on the operator coefficients are \cite{boundsR}
\begin{equation}
\frac{C_{\Pi\Pi gg}}{f^2} < \frac{1}{(120\, \mathrm{GeV})^2},\qquad 
 \frac{C_{\eta\Pi\Pi} c_G\, }{f M_\X^2} < \frac{1}{(180\, \mathrm{GeV})^2}.
\end{equation}
Both bounds roughly imply $f \gtrsim 100\ \mathrm{GeV}$.


\begin{figure}[t]
\begin{center}
\includegraphics[width=.46\textwidth]{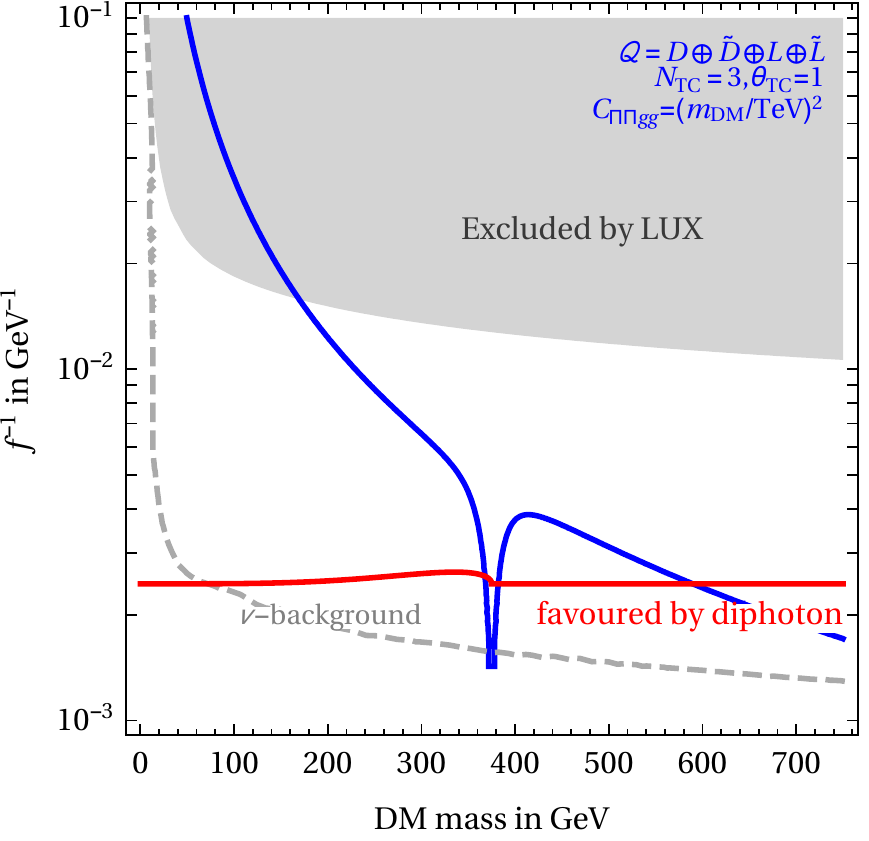}\qquad
\includegraphics[width=.46\textwidth]{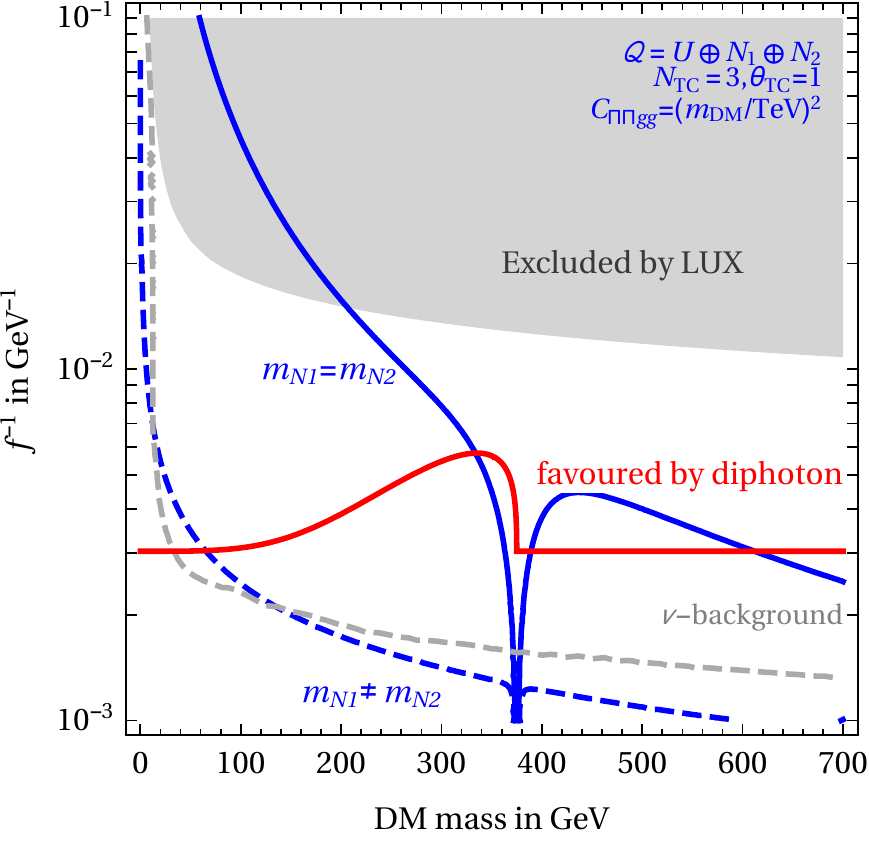}
\caption{\label{fig:UNN}\em In the left (right) panel we consider the model $\Q=D\oplus \tilde{D}\oplus L \oplus \tilde{L}$ of section \ref{sec:Baimodel} (the model $\Q=U\oplus N_1\oplus N_2$ of section \ref{sec:UNN}). Along the blue curves (solid and dashed) the Dark Matter relic density is reproduced.  
The red curves correspond to the value of $f$ that explains the di-photon rate for $N_{\rm TC}=3$, $\theta_{\rm TC}=1$.
The gray region is excluded at $90\%\,{\rm CL}$ from direct DM searches at LUX~\cite{LUX}. 
} 
\end{center}
\end{figure}

\bigskip

Figure \ref{fig:UNN} shows that it is possible to reproduce the observed DM abundance 
compatibly with direct detection constraints for values of the parameters
favoured by the 750 GeV $\gamma\gamma$ anomaly, 
although the DM mass needs to be somehow near to the resonance condition,
$m_\Pi \circa{<}M_\X/2$. However, when co-annihilations are present the DM mass is below 100 GeV as shown by the dashed blue curve in the right panel of figure \ref{fig:UNN}.
This encourages us to try to build models that realise this scenario.


\section{Confining di-photon resonance and Dark Matter }\label{4}
Our goal is constructing composite models where:
i) the 750 GeV resonance $\X$ is a composite TC$\pi$ with QCD and QED anomalies;
ii) DM is another composite TC$\pi$, $\Pi$, stable because of species number, 
iii) $\X\to \Pi\Pi^*$ is allowed.
iv) All experimental bounds are satisfied and no other TC$\pi$ is stable.
Of course, these goals go beyond what is safely indicated by experiments and  might be too ambitious.
In section~\ref{2S} we discuss the problems of models with two species.
In section~\ref{sec:UNN} we discuss models with three species, where DM is accidentally stable thanks to species number.
In section~\ref{G} we discuss models where DM is stable because of $G$ parity.

\subsection{Models with two identical species?}\label{2S}

To start, we consider models containing TCq that fill 2 identical copies of a representation $X$ of
the SM gauge group. Three kind of singlet TC$\pi$ are formed:
1) 
$\Pi=X_1 \bar{X}_2$, which is a stable DM candidate;
2) $\eta_- =X_1\bar{X}_1-X_2\bar{X}_2$, with no anomalies; and 
3) $\eta_+ = X_1\bar{X}_1+X_2\bar{X}_2$ with anomalies under the SM and under the techni-colour group.
TCq masses $m_1$ and $m_2$ contribute to the masses of the neutral states as
\beq \Delta V(\eta_\pm,\Pi) =  B_0 (m_1+m_2) \left( \Pi \Pi^* + \frac{1}{2} \eta_-^2 + \frac{1}{2} \eta_+^2 \right) + B_0 (m_1-m_2)\eta_-\eta_+   +\cdots\eeq
Furthermore, the techni-anomaly gives a large mass term to  $\eta_+$.
Because of the mixing both mass eigenstates $\eta\approx \eta_-$ (lighter) and $\eta'\approx \eta_+$ (heavier) acquire anomalous couplings with SM gauge bosons.
In the limit of large $m_{\eta'}$, $\eta$ and $\Pi$ are quasi-degenerate so that $\eta\to\Pi\Pi^*$ decays are kinematically forbidden.

In order to obtain $\X\to gg$ decays  $X$ should be coloured, leading to the following phenomenological issue.
Besides the neutral singlet $\Pi,\eta_+,\eta_-$ there are coloured TC$\pi$ (e.g.\ $3\otimes\bar 3 = 1 \oplus 8$).
A TC$\pi$ in the $(r_3,r_2)_Y$ rep acquires the following contribution to its squared mass from SM gauge interactions:
\begin{equation}
\label{Delta}
\Delta_{(r_3,r_2)_Y}
 \approx \frac{3}{4\pi}  \left( \alpha_3 C(r_3) +\alpha_2 C(r_2) +\alpha_Y Y^2 \right) \Lambda_{\rm TC}^2\,,
\end{equation}
where $C(r_{N})$ is the quadratic Casimirs of the $r_N$ representation of $\SU(N)$, 
equal to \mbox{$(N^2-1)/{2N}$} for the fundamental and to $N$ for the adjoint. 
Numerically $\Delta_{(1,3)_0}\approx 0.015 \, \Lambda_{\rm TC}^2$ for a triplet of $\SU(2)_L$,
$\Delta _{(3,1)_Y}\approx 0.03 \, \Lambda_{\rm TC}^2$ for  a colour triplet and 
$\Delta_{(8,1)_0} \approx 0.07 \, \Lambda_{\rm TC}^2$ for a colour octet.
These numerical values imply that, while the coloured TC$\pi$ become unstable (decaying to gluons and uncoloured TC$\pi$),
it seems difficult to avoid conflicting with LHC bounds that roughly excluded coloured
particles lighter than about 1 TeV (notice however that this bound is model dependent, although fairly correct for a large class of scenarios, as discussed in section \ref{8}).
Furthermore, co-annihilations between the coloured and the neutral states render difficult to reproduce the
cosmological DM thermal abundance for sub-TeV masses~\cite{1402.6287}. 

In conclusion, to build a viable model where $\X$ decays into DM we need to  add a third specie, which is heavier and coloured.

\medskip

\subsection{Dark Matter stability from species number}\label{sec:UNN}
In view of the previous considerations, we consider models with three species, as listed in  table \ref{table:UNN}.
As their phenomenology is similar, we
explicitly discuss the model with 
\begin{equation}\label{eq:UNN-embedding}
\Q =U \oplus N_1\oplus N_2 \, .
\end{equation}
The TC$\pi$ transform in the adjont representation of the techni-flavour group SU(5) that, with the above embedding, decomposes under the SM as
\begin{equation}
\label{eq:UNN-pions}  
24 = \underbrace{(8,1)_0}_{\chi 
} \oplus \underbrace{2 \times [(\bar{3},1)_{-2/3} +({3},1)_{2/3}]}_{\phi_{1,2},\phi^*_{1,2} 
} \oplus \underbrace{4 \times(1,1)_0}_{\Pi,\Pi^*, 
~\eta_{1,2}}.
\end{equation}
The TC$\pi$ with a net species numbers are the two colour triplets, $
\phi_i= U\bar{N}_i
$
and the complex singlet $\Pi=N_1 \bar{N}_2
$,
which is the DM candidate.
Stability of the triplets can  be avoided by appropriate higher dimensional operators or by
adding scalars $H'$ with quantum numbers  such that the Yukawa interactions $H'UN_i$ is allowed.
In special models such as $\Q = Q\oplus \tilde{U}_{1,2}$ or $\Q = Q\oplus \tilde{D}_{1,2}$ the role of $H$ can be played by the SM Higgs doublet~\cite{strongDM}.
The two real singlets and the octet $\chi = U \bar{U}$ are unstable and  decay through anomalies to SM gauge bosons. 
The singlets can be di-photon candidates.
Including the $\eta'$, the TC$\pi $ matrix reads
\begin{equation}
\Pi(x) = \left(\begin{array}{c|cc} \chi & \frac{\phi_1}{\sqrt{2}} & \frac{\phi_2}{\sqrt{2}} \\ \hline \frac{\phi_1^*}{\sqrt{2}} & 0 & \frac{\Pi}{\sqrt{2}}\\ \frac{\phi_2^*}{\sqrt{2}} & \frac{\Pi^*}{\sqrt{2}} & 0 \end{array}\right)+ \eta_1 T_{\eta_1} +\eta_2 T_{\eta_2} +\eta'  \frac{\One_{\NF}}{\sqrt{2 \NF}}  \,, 
\end{equation}
where the diagonal generators associated to the $\eta_i$ singlets are 
\begin{equation}\label{eq:UNN-matrix}
T_{\eta_1} = \mathrm{diag}(1,1,1,-3/2,-3/2)/\sqrt{15},\qquad T_{\eta_2} = \mathrm{diag}(0,0,0,1,-1)/2 
\end{equation}
The accidental $\SU(5)$ global symmetry  is broken by the SM gauge interactions and by the TCq mass matrix 
\begin{equation}\label{eq:massUNN}
\mathcal{M}_{\Q}=\mathrm{diag}(m_U,m_U,m_U,m_{N_1},m_{N_2}).
\end{equation}
We compute the TC$\pi$ mass matrix from eq.~\eqref{Lmass} in the appendix.
We find
\begin{equation}\label{eq:masses-UNN}
\begin{array}{ll}
m_\Pi^2 = B_0 (m_{N_1} + m_{N_2}),\qquad  & m_{\phi_1}^2 = B_0(m_U + m_{N_1}) + \Delta_\phi, \\
m_\chi^2 = 2 B_0 m_U + \Delta_\chi,  &m_{\phi_2}^2 = B_0(m_U + m_{N_2}) + \Delta_\phi, \quad 
\end{array}
\end{equation}
where $B_0$ is of order $\Lambda_{\rm TC}$ and gauge contribution $\Delta$ are given in eq.~(\ref{Delta}).
TC$\pi$ with same quantum numbers and same species number can  mix.  In particular $\eta_i$ generically mix with $\eta'$.
In the limit where $\eta'$ is much heavier, the mass matrix of $\eta_1,\eta_2$ singlets in the basis of eq.~\eqref{eq:UNN-matrix}
is given by
\begin{equation}
B_0 \left(\begin{array}{cc}
\frac{1}{5}(4 m_U + 3 m_{N_2} + 3 m_{N_1}) & \sqrt{\frac{3}{5}}  (m_{N_2} - m_{N_1})\\
\sqrt{\frac{3}{5}}  (m_{N_2} - m_{N_1}) & m_{N_1} + m_{N_2}
\end{array}\right) .
\label{eq:22matrix}
\end{equation}
The mass eigenstates are
$\eta_{m1}=\cos\theta_{12} \eta_1 - \sin\theta_{12}\eta_2$
and 
 ${\eta}_{m2}=\sin\theta_{12} \eta_1 + \cos\theta_{12}\eta_2$ where
\begin{equation}\label{eq:mixUNN}
\tan2\theta_{12} = \frac{\sqrt{15} \, (m_{N_2}-m_{N_1})}{2 m_U-m_{N_1}-m_{N_2}}.
\end{equation}
In the limit  $m_U\gg m_{N_{1,2}}$, $\eta_{m2}$ is approximately degenerate with the DM candidate due to an accidental $\SU(2)$ symmetry.
From the mixing one finds the hierarchy $m_{\eta_{m2}} < m_\Pi$, but higher order terms in the chiral expansion should also be included at this order. 
As explained in the appendix, the $\theta_{\rm TC}$-angle modifies the mass spectrum. In the limit of small $\theta_{\rm TC}$, it is a second order effect. More interesting for our discussion is the fact that $\theta_{\rm TC}$ induces cubic couplings between techni-pions, as discussed in section \ref{sec:CPviolation}.

\medskip
In the limit $m_{N_1}=m_{N_2}$ $\eta_{m2}$ and $\Pi$ become degenerate and stable with common mass $2B_0m_{N_{1,2}}$. 
They form a triplet, $T^a$, under a global $\SU(2)$ symmetry that rotates $N_1$ and $N_2$ so that DM has 3 scalar components.
The di-photon resonance is then identified with  $\eta_1$ or $\eta'$.

\subsubsection*{Techni-pion interactions with SM vectors}
Using eq.~\eqref{c-anomalies} we compute the anomaly coefficient in the interaction basis.
The colour octet $\chi$ decays dominantly to $gg$ as well as into $\gamma g, Z g$, as already discussed in section~\ref{8}.
The anomaly coefficients for the singlets $\eta_1$ and $\eta'$ are collected in table~\ref{table:UNN} for a sample of models. 
The combination corresponding to $\eta_2$ has no anomalies because in the limit $m_{N_1} = m_{N_2}$ it becomes stable.
In presence of two possible di-photon candidates, we need to check the experimental bound presented in section \ref{extraeta}. 
In models highlighted in red  (blue) only the $\eta_1$ ($\eta'$) singlet is a viable di-photon candidate.

\medskip

For $m_{N_1}\neq m_{N_2}$, the mass eigenstate $\eta_{m2}$ inherits anomalous couplings from the mixing with $\eta_1$ and $\eta'$. 
The lighter $\eta_{m2}$ has anomaly coefficients equal to those of $\eta_{m1}$, but suppressed by $\tan\theta_{12}$,
and thereby is compatible with data for small enough mixing $\theta_{12}$, see section~\ref{extraeta}.
The signal rate is
\begin{equation}
\frac{\sigma(pp\to \eta_{m2} \to \gamma\gamma) }{ \sigma(pp\to \eta_{m1} \to \gamma\gamma)}=
\frac{\tan^2\theta_{12}}{1 - \mathrm{BR}(\eta_{m1} \to \mathrm{TC}\pi)}
\frac{C_{gg}(m_{\eta_{m2}})}{C_{gg}(m_{\eta_{m1}})}
\end{equation}
where we allowed for a branching ratio of $\eta_{m1}$ to lighter TC$\pi$ to which we now turn.

\begin{table}[t]
\begin{center}
\begin{small}
\begin{tabular}{|c||c|c|c||cccc|cc|}
\hline
$U \oplus N_1 \oplus N_2$ & $\frac{c_B}{\NTC}$ &  $\frac{c_W}{\NTC}$ &$\frac{c_G}{\NTC}$ & $\frac{\Gamma_{\gamma Z}}{\Gamma_{\gamma\gamma}}$ & $\frac{\Gamma_{Z Z}}{\Gamma_{\gamma\gamma}}$ & $\frac{\Gamma_{WW}}{\Gamma_{\gamma\gamma}}$ & $\frac{\Gamma_{gg}}{\Gamma_{\gamma\gamma}}$ & $\Gamma_\X(\!\GeV)$ & $\frac{f(\!\GeV)}{\NTC}$ \,\\  \hline
\rowcolor[cmyk]{0,0.1,0,0} $\eta'$     & $\frac{4\sqrt{2/5}}{3}$  &  0   & $\frac 1 {\sqrt{10}}$ & 0.57 & 0.082 & 0 & 180 & $-$ 
& $-$ 
\\ 
\rowcolor[cmyk]{0,0.1,0,0} $\eta_1$   & $\frac 8 {3\sqrt{15}}$  & 0 & $\frac 1 {\sqrt{15}}$ & 0.57 & 0.082 & 0 & 180 & 2.7 & 47 \\ 
\hline\hline
$D \oplus E_1 \oplus E_2$ &  $\frac{c_B}{\NTC}$ &  $\frac{c_W}{\NTC}$ & $\frac{c_G}{\NTC} $  & $\frac{\Gamma_{\gamma Z}}{\Gamma_{\gamma\gamma}}$ & $\frac{\Gamma_{Z Z}}{\Gamma_{\gamma\gamma}}$ & $\frac{\Gamma_{WW}}{\Gamma_{\gamma\gamma}}$ & $\frac{\Gamma_{gg}}{\Gamma_{\gamma\gamma}}$ & $\Gamma_\X(\!\GeV)$ & $\frac{f(\!\GeV)}{\NTC}$ \,
\\  \hline
\rowcolor[cmyk]{0,0.1,0,0} $\eta'$	  & $\frac{7}{3} \sqrt{\frac{2}{5}}$  & 0 & $\frac{1}{\sqrt{10}}$ & 0.57 & 0.082 & 0 & 60 & $-$ 
& $-$ 
\\ 
\rowcolor[cmyk]{0,0.1,0,0} $\eta_1$    & $-\frac{16}{3 \sqrt{15}}$ & 0 & $\frac{1}{\sqrt{15}}$ & 0.57 & 0.082  & 0 & 46 & 8.0 & 51 \\ 
 \hline\hline
$U \oplus E_1 \oplus E_2$ &  $\frac{c_B}{\NTC}$ &  $\frac{c_W}{\NTC}$ & $\frac{c_G}{\NTC} $  & $\frac{\Gamma_{\gamma Z}}{\Gamma_{\gamma\gamma}}$ & $\frac{\Gamma_{Z Z}}{\Gamma_{\gamma\gamma}}$ & $\frac{\Gamma_{WW}}{\Gamma_{\gamma\gamma}}$ & $\frac{\Gamma_{gg}}{\Gamma_{\gamma\gamma}}$ & $\Gamma_\X(\!\GeV)$ &  $\frac{f(\!\GeV)}{\NTC}$ \,\\  \hline
\rowcolor[cmyk]{0,0.1,0,0} $\eta'$	  & $\frac{2 \sqrt{10}}{3}$  & 0 & $\frac{1}{\sqrt{10}}$ & 0.57 & 0.082 & 0 & 29 & $-$ 
& $-$ 
\\ 
\rowcolor[cmyk]{0,0.1,0,0} $\eta_1$    & $-\frac{2}{3} \sqrt{\frac 5 3}$ & 0 & $\frac{1}{\sqrt{15}}$ & 0.57 & 0.082  & 0 & 118 & 3.7 & 49 \\ 
 \hline\hline
$Q \oplus \tilde{D}_1 \oplus \tilde{D}_2$ &  $\frac{c_B}{\NTC}$ &  $\frac{c_W}{\NTC}$ & $\frac{c_G}{\NTC} $  & $\frac{\Gamma_{\gamma Z}}{\Gamma_{\gamma\gamma}}$ & $\frac{\Gamma_{Z Z}}{\Gamma_{\gamma\gamma}}$ & $\frac{\Gamma_{WW}}{\Gamma_{\gamma\gamma}}$ & $\frac{\Gamma_{gg}}{\Gamma_{\gamma\gamma}}$ & $\Gamma_\X(\!\GeV)$ &  $\frac{f(\!\GeV)}{\NTC}$ \,\\  \hline
\rowcolor[cmyk]{0.1,0,0,0} $\eta'$	  & $\frac{5}{6 \sqrt{6}}$  & $\frac{1}{2} \sqrt{\frac{3}{2}}$ & $\sqrt{\frac{2}{3}}$ & 1.8 & 4.7 & 15 & 963 & 1.7 & 110 \\ 
\rowcolor[cmyk]{0.1,0,0,0} $\eta_1$    & $-\frac{1}{2 \sqrt{6}}$ & $\frac{1}{2} \sqrt{\frac{3}{2}}$ & $0$ & 17 & 22  & 79 & 0 & $-$ & $-$ \\ 
 \hline\hline
$Q \oplus \tilde{U}_1 \oplus \tilde{U}_2$ &  $\frac{c_B}{\NTC}$ &  $\frac{c_W}{\NTC}$ & $\frac{c_G}{\NTC} $  & $\frac{\Gamma_{\gamma Z}}{\Gamma_{\gamma\gamma}}$ & $\frac{\Gamma_{Z Z}}{\Gamma_{\gamma\gamma}}$ & $\frac{\Gamma_{WW}}{\Gamma_{\gamma\gamma}}$ & $\frac{\Gamma_{gg}}{\Gamma_{\gamma\gamma}}$ & $\Gamma_\X(\!\GeV)$ &  $\frac{f(\!\GeV)}{\NTC}$ \,\\  \hline
\rowcolor[cmyk]{0.1,0,0,0} $\eta'$	  & $\frac{17}{6 \sqrt{6}}$  & $\frac{1}{2} \sqrt{\frac{3}{2}}$ & $\sqrt{\frac{2}{3}}$ & 0.15 & 1.7 & 4.2 & 279 & 2.1 &140 \\ 
\rowcolor[cmyk]{0.1,0,0,0} $\eta_1$    & $-\frac{5}{2 \sqrt{6}}$ & $\frac{1}{2} \sqrt{\frac{3}{2}}$ & $0$ & 32 & 17  & 79 & 0 & $-$ & $-$ \\ 
\hline
\end{tabular}
\end{small}
\end{center}
\caption{\em
Anomaly coefficients for the $\eta'$ and $\eta_1$  singlets and their decay widths in various models, computed in the interaction basis. In red (blue) models, only the $\eta_1$ ($\eta'$) singlet can be identified with the $750\GeV$ resonance.
The last two columns show the maximum value of the total width $\Gamma_\X$ allowed by extra decays into DM and the corresponding minimal $f$,
computed following eq.~\eqref{eq:widthbound} and \eqref{eq:moverf}. The mixing between the singlets can affect these conclusions.
\label{table:UNN}}
\end{table}

\subsubsection*{CP-violating interactions among techni-pions}
Given that TC$\pi$ are pseudo-scalars, cubic interactions among them are possible if
the $\theta_{\rm TC}$-term of techni-strong interaction violates CP.
Using the formalism described in the appendix, from eq.~\eqref{Ltrilinear-app} we find the following cubic terms $\frac{1}{2} (C_{\eta X X} \,\eta+C_{\eta' X X} \,\eta') \,X^2$ in the interaction basis:
\begin{equation}
\label{eq:cubic-UNN-1}
C_{\eta_1 \Pi \Pi^*}= C_{\eta_1 \eta_2 \eta_2} =  -\frac{a \bar{\theta}_{\rm TC}}{N_{\rm TC} f}\sqrt{\frac{3}{5}} ,\qquad
C_{\eta' \eta_1\eta_1} = C_{\eta' \Pi \Pi^*} = C_{\eta' \eta_2\eta_2}=  \frac{a \bar{\theta}_{\rm TC}}{N_{\rm TC} f}\sqrt{\frac{2}{5}}  \,,
\end{equation}
as well as $C_{\eta_2 \Pi \Pi^*}= C_{\eta_2 \eta_1 \eta_1}=0$ where  $a$ is a non-perturbative constant of order \mbox{$\sim\Lambda_{\rm TC}^2$}.
Taking into account mixing effects  \eqref{eq:mixUNN}, the decay widths for the kinematically allowed processes become 
\begin{eqnsystem}{sys:widths}
&&\frac{{\Gamma_{\eta_{m1} \to \Pi\Pi^*}}}{c^2_{12}\sqrt{1-4 m_\Pi^2/m_{\eta_{m1}}^2}}=
2 \frac{ \Gamma_{\eta_{m1} \to \eta_{m2}\eta_{m2}}}{c_{12}^2(1-\frac 8 3 s_{12}^2)^2 \sqrt{1-4m_{\eta_{m2}}^2/m_{\eta_{m1}}^2}}
=   
  \frac{3}{80\pi} \,\frac{a^2 \bar\theta^2_{\rm TC}}{N_{\rm TC}^2 f^2 m_{\eta_{m1}}} \, ,\\
&&\frac{\Gamma_{\eta' \to \Pi\Pi^*}}{\sqrt{1-4 m_\Pi^2/m_{\eta'}^2}}  = 2 \,  \frac{\Gamma_{\eta' \to \eta_{m1} \eta_{m1}}}{\sqrt{1-{4 m_{\eta_{m1}}^2}/{m_{\eta'}^2}}}  = 2 \, \frac{\Gamma_{\eta' \to \eta_{m2}\eta_{m2}}}{\sqrt{1-{4 m_{\eta_{m2}}^2}/{m_{\eta'}^2}}}  =  \frac{1}{40\pi}\frac{a^2 \bar\theta^2_{\rm TC}}{N_{\rm TC}^2 f^2 m_{\eta'}}  ,
\end{eqnsystem}
where $s_{12}=\sin\theta_{12}$ and $c_{12}=\cos\theta_{12}$.

\medskip

The parameter $\bar\theta_{\rm TC}$ is  determined by the $\theta_{\rm TC}$ angle and by the TC$\pi$ masses, as dictated by the Dashen equations \eqref{dashen}.
$\bar\theta_{\rm TC}$ is small when $\theta_{\rm TC}$ is small or any TCq is much lighter than $\Lambda_{\rm TC}$.
A  simple result for the reference model $U \oplus N_1 \oplus N_2$ is obtained in the limit where the $\eta'$ is heavy and $m_U \gg m_{N1} \approx m_{N2}$:
\be
\label{eq:dashen-limit}
\frac{a \, \bar\theta_{\rm TC}}{\NTC} \sim  m_{\Pi,\eta_2}^2 \, \tan\left(\frac{\theta_{\rm TC}}{2} \right) \,,
\ee
where $m_{\Pi,\eta_2}^2$ is the mass of the lightest almost degenerate TC$\pi$ and the formula is valid for $\theta_{\rm TC} \lesssim 1$.
The decay rate of the 750 GeV di-photon candidate $\eta_{m1}$ into DM is
\begin{equation}
\Gamma_{\eta_{m1} \to \Pi\Pi^*}\approx  1 \GeV \left(\frac{3\theta_{\rm TC}}{\NTC}\right)^2 r^4 \sqrt{1-r^2},\qquad
r \equiv \frac{2m_\Pi}{m_{\eta_{m1}}}<1
\end{equation}
where we chose 
$f\simeq 100\ \mathrm{GeV} \NTC$ to match the di-photon rate, a small mixing $\theta_{12}\ll 1$ and the limiting case of eq.~\eqref{eq:dashen-limit}.
The maximal width is obtained for $r\approx 0.90$, but still it is more than one order of magnitude below the  width favoured by ATLAS.
We also checked that adding a larger number of light singlets TCq does not help in achieving a larger width.
The difficulty in getting a large width from CP-violating decays to DM can be understood from the Dashen equations, eq.~\eqref{dashen}. 
They imply $|a \bar{\theta}_{\rm TC}/\NTC| < \mathrm{min}_i {|2 B_0 m_i|}$, therefore when one TCq becomes light the size of CP-violation diminishes. 
This is the region where a DM candidate is lighter than $M_\X/2$.

\medskip

Furthermore, $\eta_{m1}$ can  decay into $\eta_{m2}\eta_{m2}$, which, in turn, 
decays to SM gauge bosons thanks to the anomaly acquired via its mixing with $\eta_1$, 
giving rise to a final state with 4 SM vectors.
The rate of this process is $\Gamma_{\eta_{m1}\to \eta_{m2}\eta_{m2}}/\Gamma_{\gamma\gamma}$ times the di-photon rate, assuming a dominant branching ratios to di-jet for $\eta_{m2}$. Searches for pairs of di-jets set a limit of $2\div 3\ \mathrm{pb}$ at $8 \TeV$ for pair produced di-jet resonances with mass  $\approx 300$ GeV \cite{Khachatryan:2014lpa}. Imposing the di-photon constraint and rescaling to 13 TeV, we get the limit ${\Gamma_{\eta_{m1}\to \eta_{m2}\eta_{m2}}} \lesssim 10^3 \, {\Gamma_{\gamma\gamma}}$. In the present scenario this constraint is satisfied since
\begin{equation}
\frac{\Gamma_{\eta_{m1}\to \eta_{m2}\eta_{m2}}}{\Gamma_{\gamma\gamma}}\approx\ 200\, \left(\frac{3\theta_{\rm TC}}{\NTC}\right)^2 \qquad \mathrm{for}\,\, m_{\eta_{m2}}\approx 300\, \mathrm{GeV}.
\end{equation}
For smaller $m_{\eta_{m2}}$  the limits degrade quickly. We can also have final states with photons, but they are suppressed at the level of $\approx 10^{-2}\ \mathrm{fb}$ for $jj\gamma\gamma$ at 13 TeV, due to the relative suppression $\Gamma_{\gamma\gamma}/\Gamma_{gg}$ as from table \ref{table:UNN}.


\subsubsection*{Regimes for Dark Matter in models with species number}

With the interactions derived above  two different regimes for DM can be realised in this model. 
For $m_{N_1}\ne m_{N_2}$, from the mass diagonalization there is always a state lighter than DM, $\eta_{m2}$, decaying into SM gauge bosons. 
Annihilation induced by  $\Pi \Pi^* \to \eta_{m2}\eta_{m2}$  scattering easily dominates in the regime $m_{\Pi}>m_{\eta_{m2}}$. 
This scenario is illustrated in the right panel of figure \ref{fig:UNN}. The dashed blue curve reproducing the observed relic abundance 
is consistent with the required di-photon rate for DM masses below about $100\ \mathrm{GeV}$\footnote{In the present case the cross section times velocity for the TC$\pi$ scattering $\Pi \Pi^* \to \eta_{m2} \eta_{m2}$ is
\begin{equation}
\sigma \, v=
\sqrt{1-\frac{4m_{\eta_{m2}}^2}{s}}  \, 
\frac{\bigg(\left(s-\frac{2m_{\eta_{m2}}^2 +m_\Pi^2}{3}\right)
c_{12}^2+3m_\Pi^2s_{12}^2/5
+2s_{12}c_{12}\left(2 B_0 m_{N_2}- m_\Pi^2\right)/\sqrt{15}
\bigg)^2}{32\pi\, s f^4},
\end{equation}
Notice that everything can be expressed in terms of $m_\Pi$ and $m_{\eta_{m2}}$ (and the mass of the di-photon resonance).
}.
The tree-level co-annihilations dominate over other interactions and the relic abundance is reproduced for $m_\Pi \sim 50\GeV$. 

In the limit  $m_{N_1}=m_{N_2}$
$\Pi$, $\Pi^*$ and $\eta_{m2}$ form a degenerate triplet, and $\eta_{m2}=\eta_2$ becomes 
an extra stable DM candidate due to the enhanced SU(2) symmetry of the fundamental Lagrangian. 
This case is depicted in the right panel of figure \ref{fig:UNN} (solid blue curve):
$\Pi$ and $\eta_{m2}$ have in this case the correct thermal abundance for masses close to $M_\X/2$,
and the annihilation cross section is  mainly determined by CP-violating interactions. 

\bigskip 
\bigskip

To conclude let us discuss the main differences between the $\Q=U\oplus N_1\oplus N_2$ model considered so far and
models such as $\Q=U\oplus E_1\oplus E_2$, where $E_{1,2}$ are charged.
$\Pi \sim E_1\bar E_2$ is again a neutral state, candidate to be Dark Matter.
The electro-magnetic anomaly needed to achieve $\X\to\gamma\gamma$ receives extra contributions from $E_{1,2}$.
Furthermore, $\Gamma(\X\to gg)$ can be reduced  by assuming that $U$, the colored TCq,
has a mass $m_U$ above the confinement scale: in such a case only the TC$\pi$ made of $E_{1,2}$ remain light.
As discussed in section~\ref{GammaS} this allows to reproduce the di-photon excess with a larger $\Gamma_\X$.
Another difference concerns techni-baryons: in the $U\oplus N_{1,2}$ models the stable lightest techni-baryon can be neutral state, being made of $N_i$,
while this does not happen in models where $N_i$ are replaced by charged states.

\subsection{Dark Matter stability from $G$-parity}\label{G}
\label{sec:Baimodel}
We  re-analyse the model presented in~\cite{bai}.
In our notation it corresponds to the choice  $\Q=D\oplus\tilde{D}\oplus L\oplus\tilde{L}$, which allows to  impose
a generalised $G$-parity symmetry that exchanges $L\leftrightarrow \tilde L$ and $D\leftrightarrow \tilde D$.
This implies  $m_D=m_{\tilde{D}}$ and $m_L=m_{\tilde{L}}$ and that
techni-pions are classified as even or odd under this $G$-parity: 
the lightest $G$-odd techni-pion is stable (see \cite{1005.0008} for the first discussion of techni-pion DM with $G$-parity).

\smallskip

The model has $\SU(10)$ techni-flavour symmetry.
The SU(5) generators in SU(10) are $T^a = \mathrm{diag}(t^a,-(t^a)^*)$ where $t^a$ are in the fundamental of SU(5). 
One can define a $G$-parity transformation that combines charge-conjugation and a rotation $R=\exp(i \pi \mathcal{J})$  where $\mathcal{J}=i \tau_2 \otimes I_5$, that acts on the 10 TCq. The gauge interactions are $G$-parity invariant since $R^\dag t^a R = - (t^a)^*$.
However this $G$-parity is not an accidental symmetry: one has to impose that TCq masses respect it:
\begin{equation}
\mathcal{M}_\Q= I_2 \otimes \mathrm{diag}(m_D, m_D,m_D, m_L,m_L) \,.
\end{equation}
The 99 TC$\pi$ decomposes under SU(5) as
\begin{equation}
99 = 24^+ \oplus 24^-  \oplus 1^-  \oplus (10  \oplus 15 + \hbox{h.c.})
\end{equation}
where we have indicated the $G$-parity of each multiplet. 
The complex representations $r$ transform as $r \to - \bar{r}$ under $G$-parity. 
In implicit notation, we can schematically write  the TC$\pi$ matrix as
\begin{equation}
\Pi = \left(\begin{array}{c|c} 24^+ +  24^- & 10 +15 \\ \hline \overline{ 10} + \overline{ 15} & 24^+ - 24^- \end{array} \right) + 1^- + 1^+ 
\end{equation}
where the singlets corresponds to diagonal generators, in particular the $G$-even state corresponds to the $\eta'$  with generator $T_{\eta'} = \sfrac{\One_{\NF}}{\sqrt{2 \NF}}$.
With a further decomposition of $\SU(5)$ under the SM we have the  classification of TC$\pi$ in terms of SM multiplets\footnote{The standard composition is the following,
\bea
&&\hspace{2.5cm} 24 = \left(\begin{array}{c|c} (8,1)_0 & (3,2)_{-5/6} \\ \hline (\bar{3},2)_{5/6} & (1,3)_0 \end{array}\right) + (1,1)_0 \,, \notag \\
&&10_A=\left(\begin{array}{c|c} (\bar{3},1)_{-2/3} & (3,2)_{1/6} \\ \hline -(3,2)^t_{1/6} & (1,1)_1 \end{array}\right)\,, \quad 15_S=\left(\begin{array}{c|c} (6,1)_{-2/3} & (3,2)_{1/6} \\ \hline (3,2)^t_{1/6} & (1,3)_1 \end{array}\right)
\eea
}. 
The $G$-even states in the 24$^+$ are associated to the generators $T_{+}^a =\mathrm{diag}(t^a,t^a)$, 
while the stable $G$-odd $24^-$ are associated to the generators
$T_{-}^a=\mathrm{diag}(t^a,-t^a)$.
Using  eq.~\eqref{Lmass} in the appendix we compute the mass spectrum of the TC$\pi$. For charged ones,
\begin{eqnarray}
&&m_{(1,3)_{0}}^2 = 2 B_0 m_L +\Delta_{(1,3)_{0}} \,,\, m_{(8,1)_0}^2 = 2 B_0 m_D +\Delta_{(8,1)_0}\,, \, m_{(3,2)_{5/6}}^2 = B_0 (m_D + m_L) +\Delta_{(3,2)_{5/6}} \nonumber\\
&&m_{(6,1)_{2/3}}^2 = 2 B_0 m_D + \Delta_{(6,1)_{- 2/3}}\,, \, \qquad m_{(3,2)_{1/6}}^2 = B_0 (m_D + m_L)+\Delta_{(3,2)_{1/6}} \, \, \\ 
&& m_{(1,3)_{1}}^2 = 2 B_0 m_L+\Delta_{(1,3)_{1}} \,, \,
m_{(\bar{3},1)_{2/3}}^2 = 2 B_0 m_D + \Delta_{(\bar{3},1)_{- 2/3}}\,, \, 
m_{(1,1)_{1}}^2 = 2B_0 m_L + \Delta_{(1,1)_{1}}  \,.  \nonumber
\end{eqnarray}
To compute the masses of singlet TC$\pi$ we must take into account that  states with equal $G$-parity and equal quantum numbers can mix:
the $1^-$ can mix with the singlet from $24^-$,  and the $\eta'$ with the even singlet $\eta$ in the $24^+$. 
Choosing the following basis of generators
\bea
&&T_\eta = \frac{1}{2 \sqrt{30}} \, {\rm diag} (2,2,2,-3,-3,2,2,2,-3,-3) \,, \notag \\
&& T_{1^-_A} = \frac{1}{2 \sqrt{2}} \, {\rm diag} (0,0,0,1,1,0,0,0,-1,-1) \,,  \\
&&T_{1^-_B} = \frac{1}{2 \sqrt{3}} \, {\rm diag} (1,1,1,0,0,-1,-1,-1,0,0) \,,\notag
\eea
the only mixing  arises between $\eta$ and $\eta'$. The mass matrix for the singlets ($1_A^-,1_B^-,\eta,\eta'$) is block diagonal:
\begin{equation}
\left(
\begin{array}{cccc}
 2 B_0 m_L & 0 & 0 & 0 \\
 0 & 2 B_0 m_D & 0 & 0 \\
 0 & 0 & B_0(\frac{4}{5}  m_D +\frac{6}{5}  m_L) & \frac{2}{5} \sqrt{6} B_0 (m_D-m_L)\\
 0 & 0 & \frac{2}{5} \sqrt{6} B_0 (m_D-m_L) & \frac{10 a}{N_{\rm TC}}+ B_0(\frac{6}{5}  m_D +\frac{4}{5} m_L) \\
\end{array}
\right).
\end{equation}
It follows that for $m_L < m_D$ the DM candidate  $1_A^-$ can be lighter than $M_\X/2$.
The  $\eta/\eta'$ mixing
\begin{equation}
\tan(2 \,\theta_p)= -\frac{2\sqrt{6}(m_D-m_L)}{{25 \, a}/{(B_0 \NTC)} + (m_D - m_L)}
\end{equation} 
 is sizeable when $m_{\eta'}^2 \sim 10 a/N_{\rm TC}$ is comparable to the other TC$\pi$ masses. 

\subsubsection*{Interactions of the techni-pions}

The $G$-even states in real representation of the SM can decay to SM gauge bosons via anomalies. 
The anomaly coefficients for the unstable singlets $\eta$ and $\eta'$, as defined in eq.~\eqref{c-anomalies}, are given in table~\ref{table:Bai}, 
together with the ratios between the widths into $\gamma Z$, $ZZ$, $WW$, $gg$ and the width to $\gamma\gamma$. 
Following the discussion of section \ref{extraeta}, we identify the lighter $\eta$ singlet with the di-photon resonance.
Actually, because of the $\eta/\eta'$ mixing, the anomaly coefficients of the mass eigenstates are linear combinations of those reported in table~\ref{table:Bai}. 

\begin{table}[t]
\begin{center}
\begin{tabular}{|c||ccc|cccc|cc|}
\hline
$D \oplus \tilde{D} \oplus L \oplus\tilde{L}$ & $\frac{c_B}{\NTC}$ &  $\frac{c_W}{\NTC}$ &$\frac{c_G}{\NTC}$ & $\frac{\Gamma_{\gamma Z}}{\Gamma_{\gamma\gamma}}$ & $\frac{\Gamma_{Z Z}}{\Gamma_{\gamma\gamma}}$ & $\frac{\Gamma_{WW}}{\Gamma_{\gamma\gamma}}$ & $\frac{\Gamma_{gg}}{\Gamma_{\gamma\gamma}}$ & $\Gamma_\X(\!\GeV)$ & $\frac{f(\!\GeV)}{\NTC}$ \,\\  \hline
\rowcolor[cmyk]{0,0.1,0,0} $\eta'$     & $\frac{\sqrt{5}}{3}$  &  $\frac{1}{\sqrt{5}}$   & $\frac{1}{\sqrt{5}}$ & 0.23 &   1.9 & 5.0 & 180 & $-$ 
& $-$ 
\\ 
\rowcolor[cmyk]{0,0.1,0,0} $\eta$      & $-\frac{1}{3} \sqrt{\frac{5}{6}}$  & $-\sqrt{\frac{3}{10}}$ & $\sqrt{\frac{2}{15}}$ & 1.8  & 4.7 & 15 & 240 & 2.3 & 65 \\  
\hline
\end{tabular}
\end{center}
\caption{\em Anomaly coefficients for the unstable singlets $\eta'$ and $\eta$ for the model presented in section \ref{sec:Baimodel}. 
Because of the experimental bound from $\sigma(pp \to \gamma\gamma)$, only the scenario in which the singlet $\eta$ is identified with the $750$ GeV resonance is allowed.
The last columns show the maximum value of the total width $\Gamma_\X$ allowed by extra decays into DM and the corresponding minimal $f$, computed following eq.~\eq{moverf} and \eq{widthbound}. The mixing between the singlets can modify this scenario.}
\label{table:Bai}
\end{table}

\medskip

TC$\pi$ acquire CP-violating cubic interactions in the presence of the $\theta_{\rm TC}$ term.
From eq.~\eqref{Ltrilinear-app}, we can extract the cubic couplings defined as 
before eq. \eqref{eq:cubic-UNN-1},
obtaining:
\begin{equation}
\label{eq:cubic-bai}
C_{\eta{\rm TC}\pi{\rm TC}\pi} = \frac{1}{\sqrt{30}} \frac{a\bar{\theta}_{\rm TC}}{f N_{\rm TC}} \kappa,
\qquad C_{\eta'{\rm TC}\pi{\rm TC}\pi}=\frac{1}{\sqrt{5}}\frac{a\bar{\theta}_{\rm TC}}{f N_{\rm TC}}\kappa'.
\end{equation}
The relative weights in different channels are given by:
$$
\begin{small}
\begin{tabular}{c|cc|ccc|ccc|ccc}
& $1^{-}_A$ & $1^{-}_B$ & $(8,1)_0^\pm$\, & $(1,3)_0^\pm$\, &  $(3,2)_{-\frac{5}{6}}$ \,&$(6,1)_{-\frac{2}{3}} $\,  & $(3,2)_{\frac{1}{6}}$ & $(1,3)_1$ &  $(\bar{3},1)_{-\frac{2}{3}} $\,  & $(3,2)_{\frac{1}{6}}$ & $(1,1)_1$  \\ \hline
$\kappa$ & $-3$ & $2$  & $2$ & -3 &  -1/2 \,& 2\,  & $-1/2$ & $-3$ &  $2 $\,  & $-1/2$ & $-3$\\
$\kappa'$ & $1$ & $1$  & 1 & 1 &  1 \,& 1\,  & $1$ & $1$ &  $1 $\,  & $1$ & $1$ \\
\end{tabular}
\end{small}
$$

\bigskip


We can now discuss the phenomenology of the model. 
For simplicity we assume that \mbox{$m_{\eta'}\gg m_\eta$} so that the mass of the di-photon candidate is $M_\X^2=2 B_0 (\frac{2}{5}m_L + \frac{3}{5} m_D)$. This constrains the possible mass range for the two $G$-odd stable singlets $1_{A,B}^-$. Notice however that differently from the model of the previous section, the lightest TC$\pi$ in the spectrum is automatically one between $1_A^-$ and $1_B^{-}$. Defining $z=m_L/m_D$ \cite{nomura2}, the masses for the DM candidates are 
\begin{equation}
m_{1_A^-} = M_\X \sqrt{\frac{5 \, z}{3 +2\,z}},\, \qquad m_{1_B^-} = M_\X \sqrt{\frac{5}{3 +2\,z}} \,.
\end{equation}
Not all the parameter space is allowed. For large $z\circa{>}17/2$ the DM candidate ${1_B^-}$ becomes lighter than $ \sfrac{M_\X}{2}$ and
also  the coloured TC$\pi$ become lighter; in particular the mass of the colour octet is
\begin{equation}
m_{(8,1)_0} \simeq m_{1_B^-} \, \sqrt{1+ \frac{0.07 \Lambda_{\rm TC}^2}{{m_{1_B^-}}^2}}\simeq m_{1_B^-} \, \sqrt{1+ \left(\frac{265\ \mathrm{GeV}}{m_{1_B^-}}\right)^2 \NTC}
\end{equation}
where in the second step we have imposed the di-photon rate (reproduced for $f/\NTC\approx 80\ \mathrm{GeV}$), and used the relation $\Lambda_{\rm TC}\approx 4\pi f /\sqrt{N_{\rm TC}}$. For $1_B^-$ in the resonant region, we therefore expect a large $\NTC$ to comply with bounds from direct searches for coloured states.

We are then led to consider the case where $1_A^-$ is the dominant DM component and we work in the limit where the coloured states are at $\approx 1\ \mathrm{TeV}$. In this regime the interactions of $1_A^-$ with the SM are mainly mediated by the $\eta$, in particular we do not find strong constraints for the scenario where $1_A^-$ is lighter than $\frac12 M_\X$. The $\eta \to 1_A^- 1_A^-$ width is
\begin{equation}
\Gamma_{\eta\to 1_A^- 1_A^-} = \frac{3}{320}\frac{a^2 \bar\theta_{\rm TC} ^2}{\pi  f^2 M_\X N_{\rm TC}^2} \sqrt{1-\frac{4m_{1_A^-}^2}{M_\X^2}} \sim \frac{3}{320 \pi f^2} \frac{m_{1_A^-}^4}{16 M_\X}  \, \theta_{\rm TC}^2 \sqrt{1-\frac{4m_{1_A^-}^2}{M_\X^2}}
\end{equation}
where, in the last step, we used the relation $\sfrac{a \bar\theta_{\rm TC}}{\NTC} \sim  \sfrac{\theta_{\rm TC} \, m_{1_A^-}^2}{4}$ valid in the limit $m_{\eta'} \gg m_D \gg m_L$. 
The main annihilation channel is mediated by the di-photon resonance and it originates from CP violation in the composite sector, see the left panel of figure \ref{fig:UNN}. 
Co-annihilations to heavier states are negligible, the states closer in mass being $(1,1)_{1}$, which does not contribute to co-annihilation as long as $\Lambda_{\rm TC}> 5 m_{1_A^-}$, which is natural for the typical masses of the DM candidate. The other heavier stable particle $1_B^{-}$ annihilates efficiently into other (unstable) TC$\pi$ via TC$\pi$ scattering, depleting its relic density which can be roughly estimated as $\Omega_{1_B^{-}}/\Omega_{\rm DM}\sim 10^{-4} (\mathrm{TeV}/m_{1_B^{-}})^2$.

\section{Conclusions}\label{end}
A natural explanation of di-photon excess is provided by new confining gauge theories that generate singlet Nambu-Goldstone bosons 
coupled to photons and gluons through anomalies in complete analogy with pions in QCD. 
While such theories do not protect the Higgs squared mass from quadratically divergent corrections -- the Higgs and the SM particles are elementary --
they are not in tension with bounds on new physics~\cite{Kilic:2009mi} and  have been proposed in the past for various 
purposes including explaining the stability of dark matter~\cite{strongDM} and as a source for the electro-weak scale~\cite{1410.1817}.

In this note we have given a general survey of the scenarios that reproduce the di-photon excess with a composite techni-pion. 
The models under consideration are extremely predictive. Couplings to SM gauge bosons are determined by anomalies that are in turn fixed by the fermion constituents. 
The new sector should  contain new fermions that carry colour and electro-weak charges. As a consequence new resonances with SM 
quantum numbers are predicted. Coloured particles in particular will be within the reach of the LHC.
The phenomenology depends in a crucial way on the existence of a non-zero $\theta$ angle in strong sector. 
Among other effects, CP-violation can induce tree-level decays of the 750 GeV resonance $\X$ into lighter techni-pions, 
increasing the $\X$ width. We find however that these models can only reproduce a small width, at least unless the number of techni-colours is so large
that SM gauge couplings develop sub-Planckian Landau poles.

In various models such lighter techni-pions can be neutral Dark Matter candidates, stable thanks to  accidental symmetries or $G$-parity. 
Their couplings to the di-photon resonance can reproduce the observed Dark Matter relic abundance  thermally for masses around $300 \GeV$, 
while if co-annihilations are effective, masses lower than 100 GeV are favoured.

\medskip

If the di-photon excess will be confirmed, with more data from the LHC we will learn the  coupling of $\X$ to SM gauge bosons.
This will allow to infer the quantum numbers of its TCq constituents and to sharpen the possible connection with Dark Matter.
Given the simplicity and predictivity of composite models, we might soon be able to sort out the right theory.

\small 
 
\subsubsection*{Acknowledgments}
This work was supported by the ERC grant NEO-NAT.  AT is supported by an Oehme fellowship and MR by the 
MIUR-FIRB grant RBFR12H1MW. We thank Roberto Franceschini for useful discussions.

\appendix

\section{Effective Lagrangian for techni-pions}

We review the main ingredients of the effective chiral Lagrangian for TC$\pi$ (see \cite{derafael} for a comprehensive review). We focus on the explicit breaking of the techni-flavour symmetry coming from TCq masses, gauge interactions and the axial anomaly.
The NGBs are parametrised by the unitary matrix $U(x) = \exp(-2i  \Pi(x)/f)$, with
\be
\Pi(x)=\eta_1 \, \frac{\One_{\NF}}{\sqrt{2 \NF}} + \pi^a \, T^a
\ee
where $T^a$ are the generators of $\SU(\NF)$ in the fundamental representation, normalised as $\Tr(T^a \, T^b)=\frac{1}{2} \delta^{ab}$.
The effective Lagrangian in terms of the field $U$ can be written as \cite{derafael}
\bea
\label{Leff}
&&\Lag_{\rm eff}= \frac{f^2}{4} \Bigl\{ \Tr \left[ D_\mu U (D^\mu U)^\dag \right] + \Tr \left[2 B_0 \tilde{\mathcal M}_\Q (U + U^\dag)\right] +  \\
&&- \frac{a}{\NTC} \left[\bar\theta_{\rm TC}^2 - \frac{1}{4}\left(\log \left(\frac{\det U}{\det U^\dag}\right)\right)^2\right] - i \frac{a}{\NTC} \bar\theta_{\rm TC} \left[\Tr \left(U- U^\dag\right)- \log\left(\frac{\det U}{\det U^\dag}\right)\right] \Big\} + \Lag_{\rm WZW} \,. \notag
\eea
where $f$ is the TC$\pi$ decay constant, $B_0$ is a dimensional coefficient of ${\mathcal O}(\Lambda_{\rm TC})$ and $\tilde{\mathcal M}_\Q$ contains the TCq mass matrix that can be chosen diagonal. The axial anomaly induces the terms proportional to $\sfrac{a}{\NTC}$ where $a$ has dimensions of a mass squared. The factor $1/\NTC$ is expected in a large-$\NTC$ expansion \cite{veneziano} and manifestly shows that the axial anomaly disappears in the large-$N_{\rm TC}$ limit.
The parameter $\bar\theta_{\rm TC}$ is defined as
\be
\label{bartheta}
\bar\theta_{\rm TC}=\theta_{\rm TC} - \sum_j \varphi_j\,,
\ee
where $\theta_{\rm TC}$ is the analogue of the QCD $\theta$-angle and $\varphi_j$ are the phases that appear in the minimization equations of the potential energy. 
They are the solutions of the so-called Dashen equations
\be
\label{dashen}
2 \, m_i \,  B_0 \sin\varphi_i = \frac{a}{\NTC} (\theta_{\rm TC} -\sum_j \varphi_j) \qquad \qquad i,j=1,..., \NF \,,
\ee
with $m_i$ the TCq masses. Notice that $\bar\theta_{\rm TC}$ is zero if any of the TCq masses are zero. 

In eq.~\eqref{Leff} the NGBs are fluctuations around the vacuum selected by the Dashen equations. In this basis, the effects of the axial anomaly are also present in the mass matrix that can be written as $\tilde{\mathcal M}_\Q = {\rm diag} (m_i \cos\varphi_i)$. The mass terms for the NGBs can be extracted from the second and the third term of eq.~\eqref{Leff},
\be
\label{Lmass}
\Lag_{\rm mass} =  - 2 \, B_0 \, \Tr[ \mathcal{M}_\Q \Pi^2 ] - \frac{a}{\NTC} \, ({\rm Tr} \, \Pi)^2 \,.
\ee
Notice that even in the chiral limit ($m_i = 0$), the singlet $\eta_1$ acquires a mass induced by the axial anomaly $m_{\eta_1}^2 \approx \sfrac{\NF a}{\NTC}$.  
If $a/\NTC \gg m_i$, the $\eta_1$ is much heavier than the other TC$\pi$ (similarly to the QCD case) and can be decoupled. 

The axial anomaly also leads to CP-violating interactions among the techni-pions. These terms come from the last term of eq.~\eqref{Leff} 
\be
\label{Ltrilinear-app}
\Lag_{\rm cubic} = \frac{2a}{3 \NTC f} \, \bar\theta_{\rm TC}  \, \Tr[ \Pi^3 ]  \,.
\ee

\subsubsection*{Effects of $\theta_{\rm TC}$ in an explicit model}

We present some analytic formulae for the $U \oplus N_1 \oplus N_2$ model considered in section \ref{sec:UNN}. In order to study the effects induced by the $\theta_{\rm TC}$-angle on the mass spectrum and techni-pions interactions, we need to solve the Dashen equations \eqref{dashen}. For general values of the TCq masses and of $\sfrac{a}{\NTC}$, those cannot be solved analytically. 
In order to get analytic results, let us consider the limit
\be
m_{\eta'} \gg m_U \gg m_{N_1}, m_{N_2}
\ee 
that is also relevant for the phenomenology discussed in section \ref{4}.
In this limit a simple and exact solution for the Dashen equations is \cite{Witten-chiral}:
\be
\label{sol}
\frac{\sin \varphi_{N_1}}{m_{N_2}} = \frac{\sin \varphi_{N_2}}{m_{N_1}}  = \frac{\sin \theta_{\rm TC}}{\sqrt{m_{N_1}^2 + m_{N_2}^2 + 2 m_{N_1} m_{N_2} \cos \theta_{\rm TC}}}  \,, \quad \varphi_U = \mathcal{O}\left(\frac{a \bar\theta_{\rm TC}}{2 B_0 m_U \NTC}\right)  \,.
\ee
The $\theta_{\rm TC}$-angle modifies the techni-pions mass spectrum with the substitution $m_{N_i} \to m_{N_i} \cos\varphi_{N_i}$:
\begin{equation}
\begin{array}{ll}
m_\Pi^2 = B_0 (m_{N_1} \cos\varphi_{N_1} + m_{N_2} \cos\varphi_{N_2}),\qquad  & m_{\phi_1}^2 = B_0(m_U + m_{N_1} \cos\varphi_{N_1}) + \Delta_\phi, \\
m_\chi^2 = 2 B_0 m_U + \Delta_\chi,  &m_{\phi_2}^2 = B_0(m_U + m_{N_2} \cos\varphi_{N_2}) + \Delta_\phi, \quad 
\end{array}
\end{equation}
where the contributions $\Delta$ from gauge interactions are defined in eq.~(\ref{Delta}).
In the same way, the mixing (squared) mass matrix between the singlets $\eta_1$ and $\eta_2$ becomes
\begin{equation}
B_0 \left(\begin{array}{cc}
\frac{1}{5}(4 m_U + 3 m_{N_2} \cos \varphi_{N_2} + 3 m_{N_1} \cos \varphi_{N_1}) & \sqrt{\frac{3}{5}}  (m_{N_2} \cos\varphi_{N_2} - m_{N_1}\cos\varphi_{N_1})\\
\sqrt{\frac{3}{5}}  (m_{N_2} \cos\varphi_{N_2} - m_{N_1} \cos\varphi_{N_1}) & (m_{N_1} \cos\varphi_{N_1} + m_{N_2} \cos\varphi_{N_2})
\end{array}\right) .
\label{eq:22matrix}
\end{equation}

The CP-violating trilinear couplings of eq.~\eqref{eq:cubic-UNN-1} are parametrized by the $\bar \theta_{\rm TC}$ parameter, that is related to $\theta_{\rm TC}$ and to the TCq masses by the Dashen equations. The solution \eqref{sol} corresponds to 
\be
\frac{a}{\NTC} \, \bar\theta_{\rm TC} = \frac{2 B_0 m_{N_1} m_{N_2} \sin \theta_{\rm TC}}{\sqrt{m_{N_1}^2 + m_{N_2}^2 + 2 m_{N_1} m_{N_2} \cos \theta_{\rm TC}}}  \,.
\ee
There is an interesting limit. When the splitting, $\delta\equiv1-m_{N_1}/m_{N_2}$, between the two light quarks is small we have
\be
\frac{a}{\NTC}\,\bar\theta_{\rm TC} = m_{\Pi,\eta_2}^2(\theta_{\rm TC}=0) \sin \left(\frac{\theta_{\rm TC}}{2}\right) (1 + \mathcal{O}(\delta^2)) = m_{\Pi,\eta_2}^2 \tan\left( \frac{\theta_{\rm TC}}{2}\right)\, \left(1+ \mathcal{O}(\delta^2)\right)
\ee
where $ m_{\Pi,\eta_2}^2(\theta_{\rm TC}=0)=B_0 (m_{N_1}+m_{N_2}) $ is the mass squared of $\Pi$ and $\eta_2$ in the limit of vanishing $\theta_{\rm TC}$. In the approximation used they are related by $m_{\Pi,\eta_2}^2 = m_{\Pi,\eta_2}^2(\theta_{\rm TC} =0) \cos\left( \sfrac{\theta_{\rm TC}}{2} \right) \, \left( 1 + \mathcal{O}(\delta^2)\right)$. Notice that the formulae derived here are valid for $\theta_{\rm TC} \lesssim 1$, that is the relevant regime for our phenomenological discussion, and in the limit $m_U \gg m_{N_1,N_2}$ and $\delta\ll 1$. In this limit the mass of the di-photon candidate is not sensitive to the $\theta_{\rm TC}$-angle, provided $m_U\gg m_{N_1,N_2}$, while the cubic interactions can be simply expressed as functions of the physical mass $m_{\Pi,\eta_2}^2$  and the $\theta_{\rm TC}$-angle. 

We can estimate the masses of the TCq as a function of $M_\X$ and $M_{\rm DM}$. In the degenerate limit $m_{N_1}=m_{N_2}$, assuming a $\Lambda_{\rm TC}$ scale of order $1 \TeV$, we get
\be
m_{N_{1,2}} \sim 60 \GeV \, \left( \frac{M_{\rm DM}}{350 \GeV} \right)^2 \,, \quad m_U \sim 700 \GeV \, \left(1 - 0.1 \left( \frac{M_{\rm DM}}{350 \GeV} \right)^2\right) \,,
\ee
where we used as a reference point the DM mass suggested by the di-photon signal and the thermal relic abundance as shown in the right panel of  figure \ref{fig:UNN}. 
In the non degenerate limit, for a small value of the mass splitting $\delta$, we get a similar result so that for $M_{\rm DM} \sim 50 \GeV$, we can estimate $m_{N_1} \sim m_{N_2} \sim  {\rm few} \GeV$ and $m_U \sim 700 \GeV$.

\bigskip

\end{document}